\documentclass[a4paper,11pt]{article}
\pdfoutput=1 

\usepackage{jheppub} 

\usepackage[T1]{fontenc} 

\usepackage{enumerate}
\usepackage{float}
\usepackage{color}
\usepackage{slashed}
\usepackage{multirow}

\usepackage{chngcntr}
\usepackage{amssymb, amsmath,mathrsfs}

\usepackage{graphics}
\usepackage{graphicx}
\usepackage{epsf}
\usepackage{epsfig}
\usepackage{float}
\usepackage{makecell}
\usepackage{multirow}
\usepackage{color}
\usepackage{xcolor}
\usepackage{subcaption}
\usepackage{tikz-feynman}

\usepackage[utf8]{inputenc}
\usepackage{amsmath}
\usepackage{amssymb}
\usepackage[normalem]{ulem}

\usepackage{mathtools}
\usepackage{amsmath}
\usepackage{adjustbox}

\newcommand\undermat[2]{
  \makebox[0.5pt][l]{$\smash{\underbrace{\phantom{%
    \begin{matrix}#2\end{matrix}}}_{ \let\scriptstyle\textstyle\text{\large $#1$}}}$}#2}
\newcommand\overmat[2]{
  \makebox[-1pt][l]{$\smash{\overbrace{\phantom{%
    \begin{matrix}#2\end{matrix}}}^{ \let\scriptstyle\textstyle\text{\large $#1$}}}$}#2}    
\usepackage{natbib}
\usepackage{tikz}
\usepackage[vcentermath]{youngtab}
\usepackage{slashed} 
\usepackage[font=small]{caption} 
\usepackage{times} 

\usepackage{bm}       
\usepackage{bbm} 

\usepackage{relsize}  

\usepackage{autobreak}  

\usepackage{makeidx} 
\usepackage{bbding} 
\usepackage{listings} 
\usepackage{ytableau}

\newcolumntype{M}[1]{>{\centering\arraybackslash}m{#1}}
\newcolumntype{N}{@{}m{0pt}@{}}

\newcommand{\ov}[1]{\overline{#1}}
\newcommand{\mc}[1]{\mathcal{#1}}

\newcommand{\comments}[1]{}

\newcommand{\beq}{\begin {equation}}  
\newcommand{\eeq}{\end   {equation}} 
\newcommand{\bea}{\begin {eqnarray}} 
\newcommand{\eea}{\end   {eqnarray}}  
\newcommand{\baa}{\begin {array}   } 
\newcommand{\eaa}{\end   {array}   }     
\newcommand{\bit}{\begin {itemize} }
\newcommand{\eit}{\end   {itemize} }
\newcommand{\be }{\begin {equation}} 
\newcommand{\ee }{\end   {equation}}
\newcommand{\nn }{\nonumber        }

\title{\boldmath Impact of Dimension-8 SMEFT operators on Diboson Productions}

\preprint{IRMP-CP3-23-15}

\author[a]{C\'eline Degrande,}
\author[a]{Hao-Lin Li}

\affiliation[a]{Centre for Cosmology, Particle Physics and Phenomenology (CP3), Universite Catholique de Louvain}

\emailAdd{celine.degrande@uclouvain.be}
\emailAdd{haolin.li@uclouvain.be}

\abstract{We for the first time identify all the dimension-8 (dim-8) SMEFT operators that can have an interference with the SM with $E^4/\Lambda^4$ enhancement in the high energy limit for the processes $q\bar{q}\to WW/WZ$. Our results therefore explicitly show that the non-interference observed for the dimension-six does not extend to dimension-eight. We compute the contributions of those dimension-8 operators to the cross-section at the 14 TeV Large Hadron Collider and compare the results with dimension-6 (dim-6) originated corrections at the order of dim-6-SM interference and dim-6 squared. We find one (two) dim-8 operator(s) can generate amplitudes of a similar order of magnitude compared with their dim-6 squared counterparts assuming unity dimensionless Wilson coefficients. During the study, new non-interference scenarios are found due to the selection rule of angular momentum as well as strong suppression due to the symmetric initial state for the proton-proton collider.  }

\begin{document} 
\maketitle
\flushbottom

\section{Introduction}
\label{sec:intro}
Since no undisputed evidence of new physics (NP) has been found at the Large Hadron Collider (LHC) since the discovery of the Higgs, precision measurements have become a major goal for the High-Luminosity LHC. In the hypothesis where the non-observation of NP signal is due to the relatively high NP scale, the Standard Model effective field theory (SMEFT) provides a suitable framework to parameterize deviations from the SM predictions according to the decoupling theorem~\cite{Appelquist:1974tg}. In the SMEFT framework, the NP effects are encoded in the Wilson coefficients of an infinite series of higher dimensional operators that are constructed with the SM field contents and obey the SM gauge symmetries:
\begin{eqnarray}
{\cal L}_{\rm SMEFT}={\cal L}_{\rm SM}+\sum_{d,i}\frac{C^{d}_i}{\Lambda^{d-4}}{\cal O}^{(d)}_i,
\end{eqnarray}
where $\Lambda$ is the characteristic NP scale, $d$ corresponds to the mass dimension of the operator starting from 5. By simple dimensional analysis, one insertion of dimension-$d$ SMEFT operators in the Feynman diagram will suppress the amplitudes by $(E/\Lambda)^{(d-4)}$, with $E$ an energy scale of the considered process. Therefore, below $\Lambda$, the leading NP effects are expected to start from the lowest dimensional operators. 
Complete and independent sets of dimension-5 (dim-5) and dimension-6 (dim-6) operators are found in the Ref.~\cite{Weinberg:1979sa} and~\cite{Grzadkowski:2010es}. Recently, the complete and independent set of dimension-8 (dim-8) operators has been found by the traditional method~\cite{Murphy:2020rsh} and by the systematic Young tableau method~\cite{Li:2020gnx}. Therefore in the SMEFT framework, the amplitude of any process can be expanded in the following form:
\begin{eqnarray}
{\cal A} = {\cal A}_{\rm SM} +\sum_{i}\frac{C^{(6)}_i}{\Lambda^2}{\cal A}_i^{(6)}
+\sum_{j}\frac{C^{(8)}_j}{\Lambda^4}{\cal A}_j^{(8)}  +\sum_{i,j}\frac{C^{(6)}_iC^{(6)}_j}{\Lambda^4}{\cal A}_{ij}^{(8)'} + \dots,
\end{eqnarray}
where we have neglected contributions from dim-5 and dim-7 operators which violate lepton or baryon numbers and are not our focus in this work.
We also explicitly factor out the dependence on the Wilson coefficients and the NP scale $\Lambda$ in front of  ${\cal A}^{(6)}$, ${\cal A}^{(8)}$ and ${\cal A}^{(8)'}$, where  ${\cal A}^{(8)'}$ represents the amplitude that contains two insertions of dim-6 operators and behaves like a dim-8 amplitude. 

However, physical observables are obtained by amplitudes squared rather than amplitudes, in this sense, we can expand  observables such as scattering cross-section with the inverse power of $\Lambda$ as well:
\begin{eqnarray}
d\sigma &&\sim |{\cal A}_{\rm SM}|^2+\sum_i\frac{C^{(6)}_i}{\Lambda^2}2\operatorname{Re}\left[{\cal A}_i^{(6)}{\cal A}^*_{\rm SM}\right]+\sum_{i,j}\frac{C^{(6)}_iC^{(6)*}_j}{\Lambda^4}{\cal A}^{(6)}_i{\cal A}^{(6)^*}_j \nonumber \\
&&+\sum_i\frac{C^{(8)}_i}{\Lambda^4}2\operatorname{Re}\left[{\cal A}_i^{(8)}{\cal A}^*_{\rm SM}\right]+\sum_{i,j}\frac{C^{(6)*}_iC^{(6)*}_j}{\Lambda^4}2\operatorname{Re}\left[{\cal A}^{(8)'}{\cal A}^*_{\rm SM}\right]+\dots.
\end{eqnarray}
From the above formula one can find that the leading contribution to the observable is coming from the interference between the dim-6 and the SM amplitudes provided that ${\cal A}_{\rm SM}$ is non-zero, then the dim-6 squared amplitude and the interference between dim-8 {or double insertion of dim-6 interactions} and SM amplitudes are at the same order in terms of expansion of inverse power of $\Lambda$. 
It is pointed out in Ref.~\cite{Azatov:2016sqh} that for a large class of 4-point amplitudes in the high energy limit, the interference between the dim-6 and SM is suppressed by the helicity selection rule originated from the supersymmetric property of the SM amplitude. 
Consequently, the next-to-leading order corrections including the interference between dim-8 and SM amplitudes are expected to have a significant impact on theoretical predictions of observables in the high energy limit. 
The reader may be confused by the simultaneous use of the high energy limit and the SMEFT framework, we clarify here that, the high energy limit should be understood as the energy regime that is large compared with the SM electroweak scale but still relatively small compared with the NP scale. 
This is closely related to the validity problem of using the EFT framework in collider physics, where the center of mass energy of the scattering process in the experiments should not be higher than the heavy NP scale $\Lambda$. At hadron colliders, this hypothesis  is hard to check especially in the high energy tails of distributions. However, the fact that such processes/tails are strongly suppressed by the parton luminosity functions helps.
In fact, in the SMEFT framework, the parton level NP effects are more prominent at higher energy, while in the meantime, theoretical predictions in such a regime suffer from a relatively large uncertainty due to the EFT expansion, so it is an art to balance these two competing effects in order to quantify the sensitivity of the LHC~\cite{Farina:2016rws}. 
The community came up with several proposals to evaluate the EFT validity problems, but no common agreement has been reached~\cite{Brivio:2022pyi}. 
In Ref.~\cite{Cohen:2021gdw}, a simplified model has been studied to demonstrate the EFT validity and the corresponding unitarity bound when taking into account the parton distribution function. The validity and consistency issues and the related dimension-8 effect of concrete NP models such as two Higgs doublet model and vector-like quark model have also been studied in Ref.~\cite{Dawson:2021xei,Dawson:2022cmu}. In this paper, we study the validity issues within the SMEFT framework and quantitatively investigate the dim-8 effects on the diboson production at the LHC. 

The dim-6 correction to the diboson production at the LHC has been extensively studied in the literature~\cite{Butter:2016cvz,Green:2016trm,Zhang:2016zsp,Baglio:2017bfe,Ellis:2018gqa,Baglio:2018bkm,Baglio:2019uty,Baglio:2020oqu,Bellan:2021dcy,Almeida:2021asy}, while the study of the effect of dim-8 operators in the SMEFT framework is still limited. For example, Refs.~\cite{Degrande:2013kka, Bellazzini:2018paj} studied the dim-8 SMEFT contribution to the process $q\bar{q}\to ZZ/Z\gamma$, and Ref.~\cite{Bellazzini:2018paj} also discussed the UV origins of the relevant dim-8 operators. 
Ref.~\cite{Azatov:2017kzw} estimates a part of the dim-8 contribution to $q\bar{q}\to WW/WZ$ and compares it with the dim-6 originated contribution, but no complete operator basis is mentioned. 
After the discovery of the complete and independent dim-8 operator basis, no comprehensive study of the dim-8 correction to the diboson production of $WW/WZ$ channel exists, therefore in this work we find out all dim-8 operators that generate contact interactions for the  production channels $q\bar{q} \to WW/WZ$ and are able to induce an enhancement of $E^4/\Lambda^4$ in the amplitude in the high energy regime. 
We find that the operator ${\cal O}_{10}$ in eq.~\eqref{eq:ops1} has a significant impact on the cross-section of $WW$ production compared with the dim-6 contribution from triple gauge boson operator ${\cal O}_{W}$, while  ${\cal O}_{12}$ and ${\cal O}_{13}$ in general have non-negligible contributions to the cross-section of $WZ$ final state.
Through the analysis of interference amplitudes, we also find new non-interference scenarios that are caused by the angular momentum selection rule and the scenario where the angular differential cross-section gets suppressed because the $q\bar{q}$ and $\bar{q}q$ initiated channels strongly cancel each other in the proton-proton (pp) collider.

Our paper is organized as follows. In section~\ref{sec:operators}, we identify all the dim-8 operators contributing contact diagrams in the quark-initiated diboson production. In section~\ref{sec:nonintclass}, we describe three different classes of non-interference that appear  in the study of SMEFT contribution to diboson production. In section~\ref{sec:result}, we present the analytical and numerical results for the dim-8 contribution to the diboson production at LHC and compare their effects to the dim-6 originated corrections. Finally, in section~\ref{sec:conclude}, we make a conclusion and point out several possible future directions following this work.

\section{The operators}\label{sec:operators}
\begin{figure}[hbt]
  \centering
  \begin{subfigure}[t]{0.49\textwidth}
    \centering
    \includegraphics[width=1\textwidth]{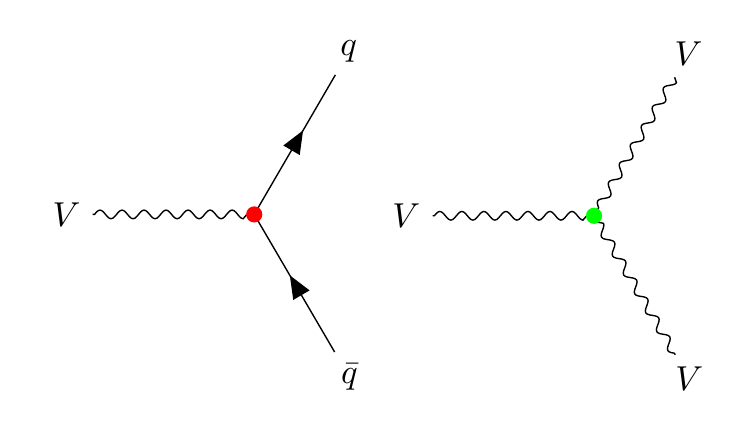}
    \caption{dim-6 vertex corrections}\label{fig:dim6vertx}
  \end{subfigure}
    \begin{subfigure}[t]{0.49\textwidth}
    \centering
    \includegraphics[width=0.59\textwidth]{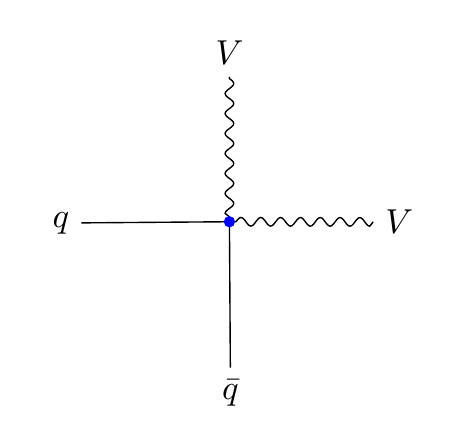}
    \caption{dim-8 contact corrections}\label{fig:dim8contact}
  \end{subfigure}
  \caption{Corrections that contribute to diboson processes from SMEFT operators.}
\end{figure}

For dim-6 SMEFT operators, corrections to the diboson production occur through their modification to triple gauge couplings (TGC) and to quark gauge interactions as shown in figure~\ref{fig:dim6vertx}. 
However, starting from dim-8, the contact interaction for $q\bar{q}\to VV$ appears as in figure~\ref{fig:dim8contact}. Moreover, among the complete and independent dim-8 operators given in Ref.~\cite{Li:2020gnx}, we find that only operators that can generate contact interactions for diboson processes can maximize the energy dependence of amplitudes to $E^4$ in the high energy regime, and  corrections to TGC and quark gauge couplings from dim-8 operators all need insertions of the Higgs vacuum expectation value (vev), which lowers the power of energy present in the amplitude by dimensional analysis. 
For this reason, we focus on the dim-8 operators that can generate contact interaction and study their interference effect with the SM amplitude in diboson production. 

Following the notation in the package ABC4EFT~\cite{Li:2022tec}, we first identify three dim-8 Lorentz classes that are capable of generating $q\bar{q}\to VV$ contact interactions ---  $F_LF_R\psi\psi^\dagger D$,  $F_LF_L\psi\psi^\dagger D$\footnote{Here we only list independent complex Lorentz classes, the Lorentz class $F_RF_R\psi\psi^\dagger D$ should be thought of as the conjugate Lorentz class of $F_LF_L\psi\psi^\dagger D$. The same concept applies to the operator level in eq.~\eqref{eq:ops2}.} and $\psi\psi^\dagger \phi^2 D^3$, where $\psi$ and $\psi^\dagger$ represent left-handed and right-handed fermion fields, $\phi$ is a scalar corresponding to the Higgs doublet or its conjugate in the SMEFT. 
$F_L$ and $F_R$ are gauge boson field strength tensors that transform as $(2,0)$ and $(0,2)$ representations of the Lorentz group, and they are related to the ordinary field strength tensor $F^{\mu\nu}$ in following formulae:
\begin{eqnarray}
F_{L/R}^{\mu\nu} = \frac{1}{2}\left(F^{\mu\nu}\mp i\tilde{F}^{\mu\nu}\right),\quad \tilde{F}^{\mu\nu}=\frac{1}{2}\epsilon^{\mu\nu\rho\sigma}F_{\rho\sigma}.
\end{eqnarray}
The reason that $\psi\psi^\dagger \phi^2 D^3$ can generate an enhancement of $E^4/\Lambda^4$ in the high energy regime can be more easily understood in the Feynman gauge, where additional derivatives acting on the Goldstones in the Higgs doublet increases the power of energy dependence in the amplitude. 
This observation also tells us that the enhancement must occur in the longitudinal-longitudinal component of the helicity amplitude as we will confirm in section~\ref{sec:result}.

After filling the $F$'s and $\psi^{(\dagger)}$'s with the concrete fields in the SM, we obtain the following independent operators for Lorentz classes $F_LF_R\psi\psi^\dagger D$, and   $F_LF_L\psi\psi^\dagger D$:
\begin{eqnarray}
&&\mc{Q}_1=iB_{R}^{\lambda\mu}B_L{}_{\lambda\nu}(d_{_\mathbb{C}}{}_p\sigma_\mu  \overleftrightarrow{D}^\nu d_{_\mathbb{C}}^\dagger{}_{r})\nonumber \\
&&\mc{Q}_2=iB_{R}^{\lambda\mu}B_L{}_{\lambda\nu}(u_{_\mathbb{C}}{}_p\sigma_\mu  \overleftrightarrow{D}^\nu u_{_\mathbb{C}}^\dagger{}_{r})\nonumber \\
&&\mc{Q}_3=iB_{R}^{\lambda\mu}B_L{}_{\lambda\nu}(Q_{p}\sigma_\mu  \overleftrightarrow{D}^\nu Q^\dagger{}_{r})\nonumber\\
&&\mc{Q}_4=i\tau^{Ii}_{j}W_{R}^{I\lambda\mu}B_L{}_{\lambda\nu}(Q_{pi}\sigma_\mu  \overleftrightarrow{D}^\nu Q^\dagger{}^{rj})\nonumber\\
&&\mc{Q}_5=i\tau^{Ii}_{j}W^{I\lambda\mu}_LB_{L\lambda\nu}(Q_{pi}\sigma_\mu  \overleftrightarrow{D}^\nu Q^\dagger{}^{rj})\nonumber\\
&&\mc{Q}_6=i W_{R}^{I\lambda\mu}W^I_{L\lambda\nu}(d_{_\mathbb{C}}{}_p\sigma_\mu  \overleftrightarrow{D}^\nu d_{_\mathbb{C}}^\dagger{}_{r})\nonumber\\
&&\mc{Q}_7=i W_{R}^{I\lambda\mu}W^I_{L\lambda\nu}(u_{_\mathbb{C}}{}_p\sigma_\mu  \overleftrightarrow{D}^\nu u_{_\mathbb{C}}^\dagger{}_{r})\nonumber\\
&&\mc{Q}_8=i W_{R}^{I\lambda\mu}W^I_{L\lambda\nu}(Q_{p}\sigma_\mu \overleftrightarrow{D}^\nu Q^\dagger{}^{r})\nonumber\\
&&\mc{Q}_9=i \epsilon^{IJK}\tau^{Ki}_{j}W_{R}^{I\lambda\mu}W^J_{L\lambda\nu}(Q_{pi}\sigma_\mu  \overleftrightarrow{D}^\nu Q^\dagger{}^{rj})\nonumber\\
&&\mc{Q}_{10}=i \epsilon^{IJK}\tau^{Ki}_{j}W_{L}^{I\lambda\mu}W^J_{L\lambda\nu}(Q_{pi}\sigma_\mu  \overleftrightarrow{D}^\nu Q^\dagger{}^{rj}),
\label{eq:ops2}
\end{eqnarray}
where $W$ and $B$ are the gauge field strength tensors of $SU(2)_L$ and $U(1)_Y$ gauge group, respectively,  the double arrow on the covariant derivative is defined as $X\overleftrightarrow{D}Y= X(DY)- (DX)Y$,  $Q$ is the left-handed quark doublets, $d_{_\mathbb{C}}$ and $u_{_\mathbb{C}}$ are charge conjugate of the right-handed down-type and up-type quarks, which also transform as left-handed fermion fields. 
All these fermion fields are in the forms of 2-component Weyl spinors, and their relations to the corresponding 4-component Dirac fields are manifested in appendix~\ref{app:spinor}. In this way, one can find the transformations between the 2-component and 4-component fermion tensors in eq.~\eqref{eq:ops2} in the following formulae:  
\begin{eqnarray}
&&J^{\mu\nu}_{d,pr}\equiv i(d_{_\mathbb{C}}{}_p\sigma^\mu  \overleftrightarrow{D}^\nu d_{_\mathbb{C}}^\dagger{}_{r})=i\bar{d}_{{\rm R}p}\gamma^\mu  \overleftrightarrow{D}^\nu d_{{\rm R}r},\nonumber \\
&&J^{\mu\nu}_{u,pr}\equiv i(u_{_\mathbb{C}}{}_p\sigma^\mu  \overleftrightarrow{D}^\nu u_{_\mathbb{C}}^\dagger{}_{r})=i\bar{u}_{{\rm R}p}\gamma^\mu  \overleftrightarrow{D}^\nu u_{{\rm R}r},\nonumber \\
&&J^{\mu\nu,I }_{(Q,3),rp}\equiv i\tau^{Ii}_{j}(Q_{pi}\sigma^\mu  \overleftrightarrow{D}^\nu Q^\dagger{}^{rj}) = i\tau^{Ii}_{j}\bar{q}_{{\rm L}r}^j\gamma^\mu \overleftrightarrow{D}^\nu q_{{\rm L}pi},\nonumber \\
&&J^{\mu\nu }_{(Q,1),rp}\equiv i (Q_{pi}\sigma^\mu  \overleftrightarrow{D}^\nu Q^\dagger{}^{ri}) = i\bar{q}_{{\rm L}r}^i\gamma^\mu \overleftrightarrow{D}^\nu q_{{\rm L}pi}.\label{eq:ftensor}
\end{eqnarray}
Pay attention that for the last equation, the two fermion fields exchange their positions on the right-hand side. 
One can also verify that the conjugates of these tensor currents are equivalent to swapping the flavor indices of the fermions, i.e. $J^{\mu\nu\dagger}_{f,pr}=J^{\mu\nu}_{f,rp}$,  and for the special case $p=r$ the fermion tensor is hermitian.
In this work, for simplicity, we only focus on the diagonal flavor components  {which have the largest interference with the SM as it is not suppressed by CKM}. 

The bosonic parts of the  operators in eq.~\eqref{eq:ops2} are expressed in terms of the chiral eigenstates $F_{L/R}$ for the sake of the advantage in enumerating the independent Lorentz structures using the Young tableau method. 
However, it is better for the phenomenological purpose to convert them into the two real degrees of freedom $F$ and $\tilde{F}$, such that the construction of a  Hermitian Lagrangian is more convenient. 
To achieve this goal, the following identities involving two field strength tensors are useful: 
\begin{eqnarray}
F_{\rm{L}\mu\rho}F_{\rm{R}}{}^{\rho\nu}&&=	\frac12F_{\mu\rho}F^{\rho\nu}+\frac18F^2\delta^{\nu}_{\mu}\label{eq:FF1}\\
\epsilon^{IJK}F^{I}_{\rm{L}\mu\rho}F^{J}_{\rm{R}}{}^{\rho\nu}
	&&=\epsilon^{IJK}\frac14\left(iF^I{}^{\nu\rho}\tilde{F}^J_{\rho\mu}+iF^I_{\mu\rho}\tilde{F}^{J\rho\nu} \right),\label{eq:FF2}\\
F_{1\rm{L}\mu\rho}F_{2\rm{R}}{}^{\rho\nu}&&=\frac{1}{4}\left\{ F_{1\mu\rho}F_{2}{}^{\rho\nu}+F_{2\mu\rho}F_{1}{}^{\rho\nu}+\frac{1}{2}[(F_1\cdot F_2)-i(F_1\cdot \tilde{F}_2)]\delta^{\nu}_{\mu}\right.\nonumber\\
&&\left.-i(\tilde{F}_{1}^{\nu\rho}{F}_{2\rho\mu}+\tilde{F}_{1\mu\rho}F_2^{\rho\nu})\right\},\label{eq:FF3}\\
F_{1\rm{L}\mu\rho}F_{2\rm{L}}{}^{\rho\nu}
	&&=\frac14\left(F_{1\mu\rho}F_2{}^{\rho\nu}-F^{\nu\rho}_{1}F_{2\rho\mu}+iF_1{}^{\nu\rho}\tilde{F}_{2\rho\mu}-iF_{1\mu\rho}\tilde{F}_2{}^{\rho\nu} \right.\nonumber \\
	&&\left. -\frac{1}{2}[(F_1\cdot F_2)-i(F_1\cdot \tilde{F}_2)]\delta^{\nu}_{\mu}\right)\;,\label{eq:FF4}\\ 
	F_{1\rm{R}\mu\rho}F_{2\rm{R}}{}^{\rho\nu}&&=\frac14\left(2F_{1\mu\rho}F_2{}^{\rho\nu}-F^{\nu\rho}_{1}F_{2\rho\mu}-iF_1{}^{\nu\rho}\tilde{F}_{2\rho\mu}+iF_{1\mu\rho}\tilde{F}_2{}^{\rho\nu} \right.\nonumber \\
	&&\left.-\frac{1}{2}[(F_1\cdot F_2)+i(F_1\cdot \tilde{F}_2)]\delta^{\nu}_{\mu}\right)\;.\label{eq:FF5}
\end{eqnarray}
For example, using eq.~\eqref{eq:FF1},  the  $B^{\lambda\mu}_R B_{L\lambda\nu}$ in ${\cal Q}_{1-3}$ and $W^{I\lambda\mu}_R W^I_{L\lambda\nu}$ in ${\cal Q}_{6-8}$ yield two different substructures for two identical gauge bosons $B$ and $W^I$ respectively, while the second term on the right-hand side of eq.~\eqref{eq:FF1} proportional to $\delta_\nu^\mu$ contracts the $\sigma_\mu D^\nu$ forming an EOM of the corresponding fermions, which can be converted to the operator of other types. 

For the Lorentz class $\psi\psi^\dagger \phi^2 D^3$, we have following operators in 2-component Weyl fermion notation:
\begin{eqnarray}
&&{\cal Q}_{11} = i(u_{_\mathbb{C}p}\sigma^\mu  \overleftrightarrow{D}^\nu u_{_\mathbb{C}}^\dagger{}^{r})(D^\mu H^\dagger D^\nu H),\nonumber \\
&&{\cal Q}_{12}=i(d_{_\mathbb{C}p}\sigma^\mu  \overleftrightarrow{D}^\nu d_{_\mathbb{C}}^\dagger{}^{r})(D^\mu H^\dagger D^\nu H),\nonumber \\
&&{\cal Q}_{13}=i(Q_{p}\sigma^\mu  \overleftrightarrow{D}^\nu Q_{}^\dagger{}^{r})(D^\mu H^\dagger D^\nu H),\nonumber \\
&&{\cal Q}_{14}=i(Q_{pi}\sigma^\mu  \overleftrightarrow{D}^\nu d_{_\mathbb{C}}^\dagger{}^{rj})(D^\mu H^{\dagger i} D^\nu H_j),\nonumber \\
&&{\cal Q}_{15}=i(u_{_\mathbb{C}p}\sigma^\mu  \overleftrightarrow{D}^\nu d_{_\mathbb{C}}^\dagger{}^{r})\epsilon^{ij}(D^\mu H_{i} D^\nu H_{j}),\label{eq:ops3}
\end{eqnarray}
where the last one, ${\cal Q}_{15}$, is an operator of the complex type for which the conjugated operator counts as an independent operator.

Using identities in eq.~\eqref{eq:ftensor}-\eqref{eq:FF5} and the Fierz identity of the $SU(N)$ gauge group, the operators in eq.~\eqref{eq:ops2} and \eqref{eq:ops3}, and their complex conjugate can be converted to the following independent operators:
\begin{eqnarray}
&&\mc{O}_1=iB^{\mu }{}{}_{\nu } B^{\nu }{}{}_{\lambda } (\bar{d}_{{\rm R}p}\gamma^\lambda  \overleftrightarrow{D}_\mu d_{{\rm R}r}),\nonumber \\
&&\mc{O}_2=iB^{\mu }{}{}_{\nu } B^{\nu }{}{}_{\lambda } (\bar{u}_{{\rm R}p}\gamma^\lambda  \overleftrightarrow{D}_\mu u_{{\rm R}r}),\nonumber \\
&&\mc{O}_3=iB^{\mu }{}{}_{\nu } B^{\nu }{}{}_{\lambda } \left(\bar{q}_{{\rm L}p } \gamma^{\lambda } \overleftrightarrow{D}_{\mu } q_{{\rm L}r}\right),\nonumber\\
&&\mc{O}_4=i W^{I\mu}{}_{\lambda}B^{\nu\lambda} \left(\bar{q}^{i}_{{\rm L}p} \gamma_{\nu } \left(\tau ^I\right)_i{}^j\overleftrightarrow{D}_{\mu } q_{{\rm L}rj}\right),\nonumber\\ 
&&\mc{O}_5=i W^{I\mu}{}_{\lambda}\tilde{B}^{\nu\lambda} \left(\bar{q}^{i}_{{\rm L}p} \gamma_{\nu } \left(\tau ^I\right)_i{}^j\overleftrightarrow{D}_{\mu } q_{{\rm L}rj}\right),\nonumber\\
&&\mc{O}_6=i W^{I\nu}{}_{\lambda}B^{\mu\lambda} \left(\bar{q}^{i}_{{\rm L}p} \gamma_{\nu } \left(\tau ^I\right)_i{}^j\overleftrightarrow{D}_{\mu } q_{{\rm L}rj}\right),\nonumber\\
&&\mc{O}_7=i W^{I\nu}{}_{\lambda}\tilde{B}^{\mu\lambda} \left(\bar{q}^{i}_{{\rm L}p} \gamma_{\nu } \left(\tau ^I\right)_i{}^j\overleftrightarrow{D}_{\mu } q_{{\rm L}rj}\right),\nonumber\\
&&\mc{O}_8=iW^{I}{}^{\mu }{}{}_{\nu } W^{I}{}^{\nu }{}{}_{\lambda }(\bar{d}_{{\rm R}p}\gamma^\lambda  \overleftrightarrow{D}_\mu d_{{\rm R}r}),\nonumber\\
&&\mc{O}_9= iW^{I}{}^{\mu }{}{}_{\nu } W^{I}{}^{\nu }{}{}_{\lambda }(\bar{u}_{{\rm R}p}\gamma^\lambda  \overleftrightarrow{D}_\mu u_{{\rm R}r}),\nonumber\\
&&\mc{O}_{10}=i W^{I}{}^{\mu }{}{}_{\nu } W^{I}{}^{\nu }{}{}_{\lambda } \left(\bar{q}_{{\rm L}r } \gamma^{\lambda } \overleftrightarrow{D}_{\mu } q^{}_{{\rm L}p}\right) ,\nonumber\\
&&\mc{O}_{11}=i \epsilon ^{IJK} W^{I}{}^{\mu }{}{}_{\nu } W^{J}{}^{\nu }{}{}_{\lambda } \left(\bar{q}^{i}_{{\rm L}p} \gamma^{\lambda } \left(\tau ^K\right)_i{}^j\overleftrightarrow{D}_{\mu } q_{{\rm L}rj}\right),\nonumber \\
&&\mc{O}_{12}=i \epsilon ^{IJK} \tilde{W}^{I}{}^{\mu }{}{}_{\nu } W^{J}{}^{\nu }{}{}_{\lambda } \left(\bar{q}^{i}_{{\rm L}p} \gamma^{\lambda } \left(\tau ^K\right)_i{}^j\overleftrightarrow{D}_{\mu } q_{{\rm L}rj}\right),\nonumber\\
&&\mc{O}_{13}=i\epsilon ^{IJK} W^{I}{}^{\mu }{}{}_{\nu } \tilde{W}^{J}{}^{\nu }{}{}_{\lambda } \left(\bar{q}^{i}_{{\rm L}p} \gamma^{\lambda } \left(\tau ^K\right)_i{}^j\overleftrightarrow{D}_{\mu } q_{{\rm L}rj}\right),\nonumber \\
&&\mc{O}_{14}=i \left(\bar{u}_{{\rm R}r } \gamma^{\lambda } \overleftrightarrow{D}_{\mu } u^{}_{{\rm R}p}\right)\left(D_\lambda H^\dagger  D^{\mu}H\right),\nonumber \\
&&\mc{O}_{15}=i \left(\bar{d}_{{\rm R}r } \gamma^{\lambda } \overleftrightarrow{D}_{\mu } d^{}_{{\rm R}p}\right)\left(D_\lambda H^\dagger  D^{\mu}H\right),\nonumber \\
&&\mc{O}_{16}=i \left(\bar{q}_{{\rm L}r } \gamma^{\lambda } \overleftrightarrow{D}_{\mu } q^{}_{{\rm L}p}\right)\left(D_\lambda H^\dagger  D^{\mu}H\right),\nonumber \\
&&\mc{O}_{17}=i \left(\bar{q}_{{\rm L}p} \gamma^{\lambda }\tau ^K\overleftrightarrow{D}_{\mu } q_{{\rm L}r}\right)\left(D_\lambda H^\dagger \tau^K D^{\mu}H\right),\nonumber \\
&&\mc{O}_{18}=i(\bar{u}_{{\rm R}p}\gamma^\mu  \overleftrightarrow{D}^\nu {d}_{{\rm R}r})\epsilon^{ij}(D^\mu H_{i} D^\nu H_{j}),
\label{eq:ops1}
\end{eqnarray}
which revive the results in Ref.~\cite{Li:2020gnx}. Restricting to the flavor diagonal components we have the following relations between ${\cal Q}$ and ${\cal O}$ operators:
\begin{eqnarray}
&&\begin{pmatrix}
{\cal O}_1 ,{\cal O}_2,{\cal O}_3,{\cal O}_8,{\cal O}_9,{\cal O}_{10}
,{\cal O}_{14},{\cal O}_{15}\end{pmatrix}
=\frac12\begin{pmatrix}
{\cal Q}_1 ,{\cal Q}_2,{\cal Q}_3,{\cal Q}_6,{\cal Q}_7,{\cal Q}_8,2{\cal Q}_{11},2{\cal Q}_{12}
\end{pmatrix},\\
&&\begin{pmatrix}
{\cal O}_4 \\{\cal O}_5\\{\cal O}_6\\{\cal O}_7
\end{pmatrix}
=\begin{pmatrix}
1 & 1 & -1 & -1 \\
i & -i & i & -i \\
1 & 1 & 1 & 1 \\
i & -i & -i & i
\end{pmatrix}
\begin{pmatrix}
{\cal Q}_4 \\{\cal Q}^*_4\\{\cal Q}_5\\{\cal Q}^*_5
\end{pmatrix},\quad
\begin{pmatrix}
{\cal O}_{11} \\{\cal O}_{12}\\{\cal O}_{13}
\end{pmatrix}
=\begin{pmatrix}
0 & 1 & 1 \\
2i & i & -i \\
-2i & i & -i 
\end{pmatrix}
\begin{pmatrix}
{\cal Q}_9 \\{\cal Q}_{10} \\{\cal Q}^*_{10}
\end{pmatrix},\\
&&\begin{pmatrix}
{\cal O}_{16} \\{\cal O}_{17}
\end{pmatrix}=
\begin{pmatrix}
1 & 0\\
-1/2 & 1
\end{pmatrix}
\begin{pmatrix}
{\cal Q}_{13} \\{\cal Q}_{14}
\end{pmatrix}.
\end{eqnarray}

\section{Different classes of non-interfering effect}\label{sec:nonintclass}
The non-interference between the SM amplitude and the amplitude with one insertion of dim-6 SMEFT operator for various of 2-to-2 processes is one of our motivations for the study of the dim-8 operators. 
The original observation in Ref.~\cite{Azatov:2016sqh} is mainly based on the helicity selection rule in the massless SM, 
which states that the interference between the helicity amplitudes --- ${\cal A}^{\rm SM}_{\{h_i\}}{\cal A}^{\rm EFT*}_{\{h_i\}}$ for the 2-to-2 scattering processes involving $VVVV$, $VV\psi\psi$, $V\psi\psi\phi$ and $VV\phi\phi$ vanishes for each set of helicity configuration $\{h_i\}$ in the massless limit, where $V,\psi,\phi$ represents spin-1,1/2,0 particle respectively. 
The reference also mentioned the possible selection rule due to the conservation of the $SU(2)$ isospin, which can be in principle extended to the $SU(3)$ color gauge group. 
Such a selection rule also exists in the relevant loop calculation, which is used to account for some vanishing entries in the anomalous dimension matrix of the Renormalization group equation of the dim-6 SMEFT operators~\cite{Craig:2019wmo}. 
Furthermore, in Ref.~\cite{Jiang:2020rwz,Shu:2021qlr} more sophisticated selection rules based on the argument of the partial wave expansion and the principle of the conservation of angular momentum help to explain more vanishing entries in the anomalous dimension matrix that Ref.~\cite{Craig:2019wmo} cannot explain. 
As we will show later, such a selection rule used for loop calculation can also benefit the analysis of the non-interference in the scattering process, as the tree-level interference cross-section for certain scattering processes can be obtained by cutting the internal propagators in the relevant loop amplitudes.    

On the other hand, researchers in the community are also interested in reviving the non-interference effect. For example, by measuring the azimuthal angle distribution of the decay product of the intermediate massive particles, one can restore the interference effects from the triple gauge dim-6 operators~\cite{Panico:2017frx,Azatov:2017kzw}. 
In the meantime, a method of quantitative evaluation of the upper bound of the interfering effect that can be revived has been proposed in Ref.~\cite{Degrande:2020tno}. 
Recently, Ref.~\cite{Hwang:2023wad} proposed a new kinematic variable --- VBFhardness to help to resurrect the non-interference in dilepton final states in probing anomalous triple gauge couplings.
In this sense, it is useful to classify different types of non-interfering effects between the SM and the SMEFT amplitudes in advance, which helps people working on resurrection programmes to identify their target operators and processes. Here we classify the non-interfering effects for the 2-to-2 scattering in the following types:
\begin{itemize}
    \item \underline{Non-interference due to the selection rule of helicities.}
    
    As mentioned above, this results from the fact that certain types of 4-point amplitudes generated by the massless SM have definite helicity combinations, which mismatch those generated by one insertion of dim-6 operators. Such a selection rule is only valid in the massless limit, and mostly for 4-point amplitudes, therefore the interfering total cross-section $\sigma^{\rm int}$ can in principle be revived by the mass effect of vector bosons and quarks or higher order corrections. On the other hand, even in the massless limit, the interference effect can be resurrected by analyzing the differential cross-section $d\sigma^{\rm int}$ for the 2-to-4 process with successive decay of the unstable particle produced in the 2-to-2 processes~\cite{Panico:2017frx,Azatov:2017kzw}. 
    
    \item \underline{Non-interference due to CP symmetries.}
    
   Since some of our operators are odd under CP transformation, an amplitude with a single insertion of a vertex from those operators, $\mathcal{A}^{CP}_{d8}$, has a weak CP phase of $\pi/2$, \textit{i.e.} the amplitude for the CP conjugate process differs by a sign,
    \begin{equation}
    \mathcal{A}^{CP}_{d8}\left(a\to b\right) = 
    -\mathcal{A}^{CP}_{d8}\left(\bar a\to \bar b\right).
    \end{equation}
    On the contrary, the SM model amplitude is CP-even if we neglect its strongly suppressed Jarlskog invariant. As a result, the interference between the SM and the dim-8 amplitudes will change sign under CP transformation, and it will not contribute to CP even observables such as the $W^+W^-$ total cross-section, invariant mass distribution, etc. Since the process is C-invariant, the interference can only contribute to observables that violate parity. In $W^+W^-$, this requires involving the helicities of the gauge bosons.
    The situation is more complicated when the process is not C-invariant such as  $W^+ Z$ as a C transformation turns it into $W^- Z$. This issue could be overcome by summing up the two processes. However, at the LHC, the two processes have different cross-sections due to the C-breaking initial state (pp) which manifests itself through different luminosities of the initial partons. Nevertheless, a similar phase difference appears in the first approximation between $\mathcal{A}^{CP}_{d8}$ and the SM amplitude and can be understood in the following way.
    At the tree level, and far away from any resonance, one can relate the complex conjugate of an amplitude ${\cal A}(i\to f)$ --- ${\cal A}^*(i\to f)$ with the amplitude of the reversed process ${\cal A}(f\to i)$ due to the optical theorem at leading order. 
    On the other hand, the reversed amplitude can be related to the original one with the time reversal transformation due to its antiunitary property. According to the CPT theorem, the T transformation is equivalent to the CP transformation, so one can derive the following relation~\cite{Jacob:1959at,Tung:1985na,Degrande:2021zpv,Azatov:2019xxn}:
    \begin{eqnarray}
    {\cal A}(a\to b)=\lambda_{CP}{\cal A}^*(a\to b),
    \end{eqnarray}
    with $\lambda_{CP}=\pm 1$
    for CP-even and CP-odd amplitudes respectively\footnote{More precisely, the CP-violating amplitude can be thought of as the tree-level amplitude obtained by any odd number of insertion of CP-violating interactions in the perturbation theory.}. Therefore, if the SM amplitude is CP-preserving and the NP amplitude is CP-violating, then their interference ${\cal A}_{\rm SM}{\cal A}^*_{\rm NP}+c.c.$ is zero for any helicity combinations.  
    Such an argument is very strong as one does not require initial and final states to be CP eigenstates. This non-interference effect cannot be revived at the total cross-section level but is possible to revive at the differential cross-section level by again studying the azimuthal angle distribution of decay products. 
   In addition, the inclusion of the width effect of the massive gauge bosons or higher order corrections in the ${\cal A}_{\rm SM}$ which generates the imaginary part of the amplitude can also resurrect this non-interference effect provided that $a$ and $b$ are not CP eigenstates. 
    
    \item \underline{Non-interference due to the selection rule of Lorentz and gauge quantum numbers.}
    
    As pointed out in Ref.~\cite{Li:2022abx}, the amplitude generated by the J-basis operator in the SMEFT can be factorized into the following form\footnote{This can also be thought of as the definition of the J-basis operators --- they generate local on-shell amplitudes for the specific scattering channel of quantum number J.}:
    \begin{eqnarray}
    {\cal A}_{\rm EFT}(i\to f)\sim \langle f|O^{J,\mathbf{R}}|i\rangle\sim {\cal T}_{\rm EFT}^{J,\mathbf{R}}\langle f|J,\mathbf{R}\rangle \langle J,\mathbf{R} |i\rangle,
    \end{eqnarray}
    which only contains the generalized partial wave amplitude of the angular momentum $J$ and the gauge quantum number $\mathbf{R}$. In the above formula, we have suppressed any additional labels for the spin and gauge indices associated with the representation $J$ and $\mathbf{R}$, ${\cal T}^{J,\mathbf{R}}$ encodes the dynamic information from the operator $O^{J,\mathbf{R}}$, and $\langle f|J,\mathbf{R}\rangle$ and $\langle J,\mathbf{R} |i\rangle$ are Poincar\'e CG coefficients. Similarly, one can do the partial wave expansion for the SM amplitude:
    \begin{eqnarray}
    {\cal A}_{\rm SM}(f\to i)\sim \sum_{J,\mathbf{R}} {\cal T}_{\rm SM}^{J,\mathbf{R}} \langle f|J,\mathbf{R} \rangle \langle J,\mathbf{R} |i\rangle,
    \end{eqnarray}
    and therefore, the interference term can be written as:
    \begin{eqnarray}\label{eq:interJbasis}
    {\cal A}_{\rm SM}{\cal A}^*_{\rm EFT}\sim \sum_{J',\mathbf{R}'} {\cal T}_{\rm SM}^{J',\mathbf{R}'}{\cal T}_{\rm EFT}^{*J,\mathbf{R}} \langle J,\mathbf{R} |f\rangle\langle f|J',\mathbf{R}' \rangle \langle J',\mathbf{R}' |i\rangle \langle i|J,\mathbf{R}\rangle,
    \end{eqnarray}
    where we deliberately separate the two CG coefficients in ${\cal A}^*_{\rm EFT}$ such that the bras and kets of initial and final states are next to each other. 
    From eq.~\eqref{eq:interJbasis}, we can see that if we sum over the gauge configurations of the initial or final states, then the inner product of state vectors for the gauge space will collapse to form $\langle \mathbf{R}|\mathbf{R}'\rangle\sim \delta_{\mathbf{R}\mathbf{R}'}$, as a consequence, if SM cannot generate  the generalized partial wave amplitude of gauge quantum number $\mathbf{R}$ in ${\cal A}_{\rm SM}$, then it does not interfere with the EFT amplitude. 
    For the same reason, integrating over the kinematic configuration of the final state (corresponds to integrating over the solid angle in the 2-to-2 scattering process) will generate the inner product between eigenstates of angular momentum $\langle J|J'\rangle\sim \delta_{JJ'}$, which selects out the specific partial wave of $J'=J$ in the SM scattering amplitude. 
    If one cannot differentiate the components of a gauge multiplet in the experiment, then such a non-interference effect due to the selection rule for the gauge quantum number is not revivable, while for the case of angular momentum, the non-interference is in principle possible to be revived by analyzing the angular distribution for the 2-to-2 process unless there are strong cancellation between different processes as we will discuss in the next section.

\end{itemize}

\section{Result}\label{sec:result}
In this section, we present the analytical and numerical results for the dim-8 operators discussed in section~\ref{sec:operators} for  different diboson final states. For each final state, we will compare the hadronic cross-section of the dim-8 and SM interference to those of the dim-6 square and dim-6-SM interference. We restrict our study to a single dim-6 operator ${\cal O}_{W}$:
\begin{eqnarray}
{\cal O}_{W } = \epsilon^{IJK}W^{I\nu}_{\mu}W^{J\rho}_{\nu}W^{K\mu}_{\rho},
\end{eqnarray}
which is known to be able to generate diboson amplitudes with $E^2/\Lambda^2$ enhancement in the high energy limit but with vanishing interference amplitude with the SM counterparts.
For dim-8 operators, we summarize their possible contribution to different final states in table~\ref{tab:modes}.

\begin{table}
\begin{center}
\begin{tabular}{ |c|c| } 
 \hline
 operator & channels  \\ 
 \hline
 ${\cal O}_1$ & $\bar{d}d\to \gamma\gamma/\gamma Z/ZZ$  \\ 
 \hline
 ${\cal O}_2$ & $\bar{u}u\to \gamma\gamma/\gamma Z/ZZ$  \\ 
 \hline
 ${\cal O}_3$ &  \makecell{$\bar{u}u\to \gamma\gamma/\gamma Z/ZZ$\\ $\bar{d}d\to \gamma\gamma/\gamma Z/ZZ$}  \\ 
 \hline
 ${\cal O}_4$, ${\cal O}_5$, ${\cal O}_6$, ${\cal O}_7$ &  \makecell{$\bar{u}u\to \gamma\gamma/\gamma Z/ZZ$\\ $\bar{d}d\to \gamma\gamma/\gamma Z/ZZ$ \\$\bar{u}d\to \gamma W/WZ$}  \\ 
 \hline
 ${\cal O}_8$ &  \makecell{ $\bar{d}d\to \gamma\gamma/\gamma Z/ZZ/WW$}  \\ 
 \hline
 ${\cal O}_9$ &  \makecell{$\bar{u}u\to \gamma\gamma/\gamma Z/ZZ/WW$}  \\ 
 \hline
 ${\cal O}_{10}$ &  \makecell{$\bar{u}u\to \gamma\gamma/\gamma Z/ZZ/WW$\\ $\bar{d}d\to \gamma\gamma/\gamma Z/ZZ/WW$}  \\ 
 \hline
  ${\cal O}_{11}$, ${\cal O}_{12}$,  ${\cal O}_{13}$ &  \makecell{$\bar{u}u\to WW$\\ $\bar{d}d\to WW$\\ $\bar{u} d\to \gamma W/WZ$}  \\ 
 \hline
   ${\cal O}_{14}$ &  \makecell{$\bar{u}u\to WW/ZZ$}  \\ 
 \hline
    ${\cal O}_{15}$ &  \makecell{$\bar{d}d\to WW/ZZ$}  \\ 
 \hline
     ${\cal O}_{16}$ &  \makecell{$\bar{u}u\to WW/ZZ$\\$\bar{d}d\to WW/ZZ$}  \\ 
 \hline
      ${\cal O}_{17}$ &  \makecell{$\bar{u}u\to WW/ZZ$\\ $\bar{d}d\to WW/ZZ$\\ $\bar{u} d\to WZ$}  \\ 
 \hline
       ${\cal O}_{18}$ &  \makecell{ $\bar{u} d\to WZ$}  \\ 
 \hline
\end{tabular}\caption{Diboson modes that dim-8 operators can generates.}\label{tab:modes}
\end{center}
\end{table}

For  analytical results, we use  {\tt FeynRules}~\cite{Alloul:2013bka} to generate the {\tt FeynArts}~\cite{Hahn:2000kx} model file and then use {\tt FeynArts} and {\tt FormCalc}~\cite{Hahn:1998yk} to generate tree level quark initiated diboson production amplitudes. 
For numerical results, we use the input scheme of $\alpha, G_F, M_Z$, with their values listed below:
\begin{eqnarray}
\alpha^{-1}=127.95,\quad G_F=1.16638\times 10^{-5}\ {\rm GeV}^{-2},\quad M_Z=91.1876\ {\rm GeV}.
\end{eqnarray}
We use the Mathematica package {\tt ManeParse}~\cite{Clark:2016jgm} to read the PDF data set in the LHAPDF6 format~\cite{Buckley:2014ana}. The PDF set we use to calculate the hadronic cross-section at the 14TeV LHC is \texttt{nCTEQ15FullNuc\_1\_1}. In the following calculation, we assume that the CKM matrix is diagonal and all the light quarks $u,d,c,s$ are massless. Yukawa couplings of light quarks are also set to zero. 
For each dim-8 operator, we assume the Wilson coefficient is non-vanishing only for the flavor diagonal component of the first generation of quark. We compare our analytical results obtained by the {\tt FeynArts} and {\tt FormCalc} at both amplitude level and cross-section level with those obtained by {\tt MadGraph}~\cite{Alwall:2014hca} and find that they are in agreement within machine precision.

\subsection{$WW$ final state}
We summarize the scaling behavior of the interference amplitudes and those after integrating over the solid angle in the high energy limit in table~\ref{tab:scalingWW}, and we save the results for those coefficients in the center of mass energy expansion, e.g. $a_i, b_i, c_i$, together with the full analytical expression for the helicity amplitudes in Mathematica files in the link~\cite{ampresult}.
As one can see only ${\cal O}_{8}$, ${\cal O}_{9}$ and ${\cal O}_{11}$ cannot generate interference amplitudes that scale as $S^2$, with $S$ the square of center of mass energy of the parton system. 
${\cal O}_{11}$ is CP odd as illustrated in appendix~\ref{app:cp}. Following the discussion in section~\ref{sec:nonintclass}, the amplitude it generates is purely imaginary, multiplied by the purely real SM CP conserving amplitude\footnote{neglecting CP violation by the CKM matrix}, the resulting interference vanishes for all helicity configurations. 
The inclusion of the width effect in the propagator in the SM amplitudes does not help to revive this non-interference effect when summing over the helicity configurations of the final states, because in this case the initial and final states are CP eigenstates, and the cancellations occur between the CP conjugate helicity configurations.
However, when taking into account the decay of the unstable vector bosons, the interference between different polarizations will contribute to the full amplitude, and thus is possible to revive the non-interference effects.

\begin{figure}[h]
\centering
\includegraphics[width=0.5\textwidth]{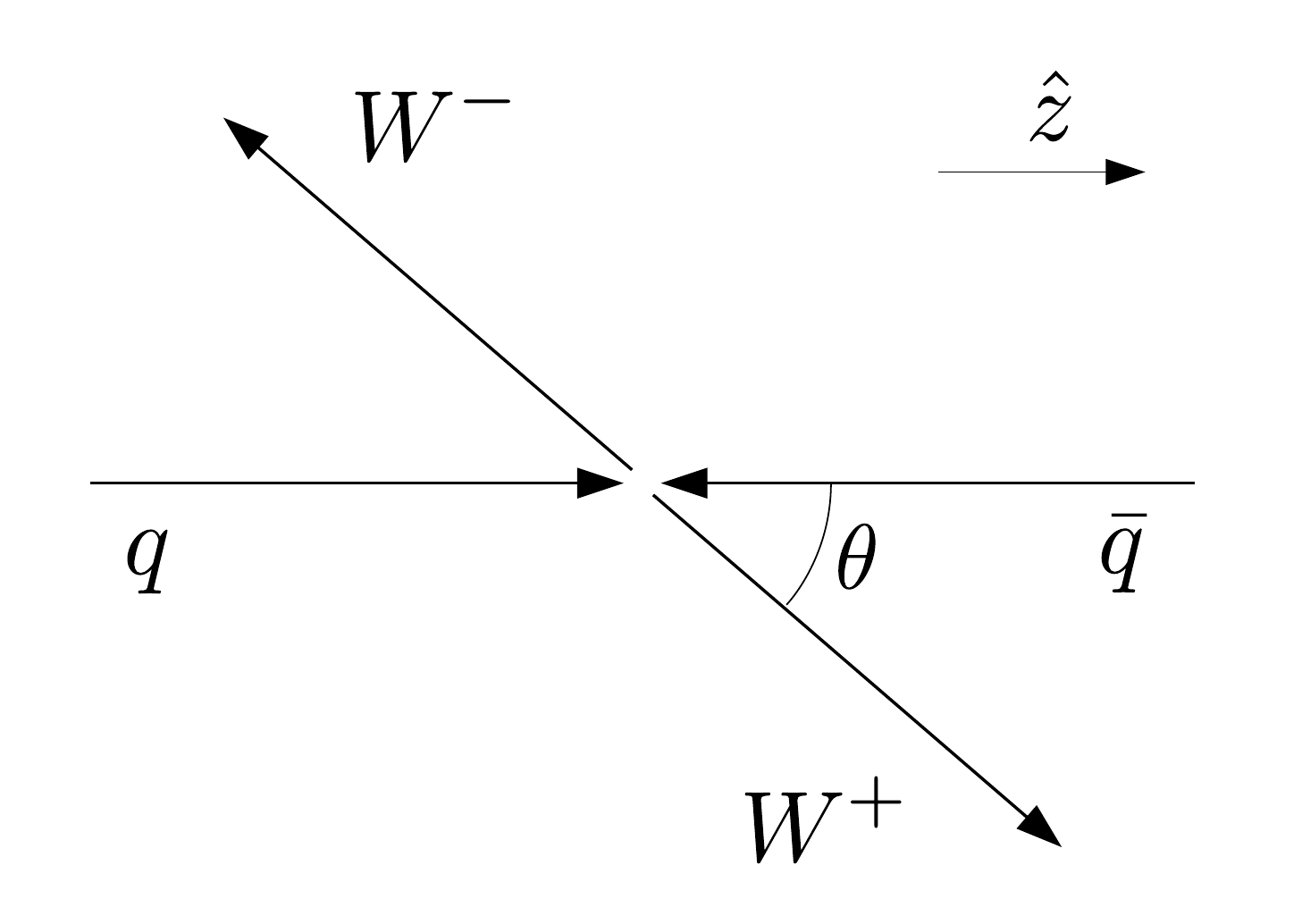}
\caption{The coordinate system in studying the $q\bar{q}\to WW$ process, $\theta$ is defined to be and angle between by the positive directions of momenta of $W^+$ and the quark $q$ in the bosons center of mass frame.}\label{fig:frame}
\end{figure}

The operator ${\cal O}_8$ and ${\cal O}_9$ in contrast generate interference amplitudes scaling linearly with $S$, and the reason can be understood with the analysis of the helicities of $W$ bosons in the final state. 
As discussed in section~\ref{sec:operators}, operators ${\cal O}_8$ and ${\cal O}_9$ are derived from ${\cal Q}_6$ and ${\cal Q}_7$ respectively, which only contain the contraction of field strength tensors of opposite chirality --- $W_LW_R$, and thus the vector bosons they produce in the high energy limit should be of opposite helicities with total angular momentum at least equal to 2. 
Meanwhile, the fermion part of the two operators only involves the right-handed up-type or down-type quarks, therefore the corresponding SM helicity amplitude only receives the contribution from the s-channel diagram with the exchange of $Z/\gamma$ boson, which cannot generate a diboson final state with a total angular momentum higher than $1$. In other words, the SM amplitude for diboson of opposite helicities is zero. 
Consequently, the non-vanishing interference  only occurs for the diboson helicity configurations $\{--,-0,+0,++,0+,0-,00\}$, which can be converted from the configurations of opposite helicities $\{+-,-+\}$ by either one insertion of the gauge boson-goldstone vertices or one insertion of gauge boson mass, and each insertion reduces the momentum dependence by respectively one or two power of the mass. 
Finally combining with the SM amplitude which at most scales as $1/\sqrt{S}$ results in the lower power dependence on $S$. We tabulate the helicity amplitudes for ${\cal O}_8$ and the SM counterparts in table~\ref{tab:O8hel}, where $\theta$ is the scattering angle between the quark and the $W^+$ in the center of mass frame of quark system as pictorially shown in figure~\ref{fig:frame}.

Similar to ${\cal O}_8$ and ${\cal O}_9$, ${\cal O}_{14}$ and ${\cal O}_{15}$ also contain the same right-handed fermion tensor structure, on the contrary, their interference amplitudes scale as $S^2$ in the high energy limit. 
The reason is that they directly produce the longitudinal modes of gauge bosons that are acted with additional derivatives --- $\partial V_L\partial V_L$, which is confirmed by the formula for the $(0,0)$ polarization in table~\ref{tab:O15hel}. Combined with the constant scaling for the SM counterpart, the overall scaling of the interference amplitude remains as $S^2$. 

Even though the interference amplitudes for  ${\cal O}_{8,9}$ and ${\cal O}_{14,15}$ do not vanish, when integrated over the solid angle they become zero, this can be explained by the third non-interference scenario we discussed in the previous section. 
Using the package ABC4EFT~\cite{Li:2022tec}, we perform the J-basis analysis for the operators, which tells us that they coincidence with the J-basis operator for the scattering channel $q\bar{q}\to WW$ with angular momentum $J=2$. 
However, as pointed out in the previous paragraph, the SM amplitude gets contribution only from the s-channel diagram of exchanging $Z/\gamma$, which corresponds to a partial wave of $J=1$. Therefore, the angular part of the NP amplitude and the corresponding SM counterpart should be proportional to the Wigner $d$-matrix of the following forms:
\begin{eqnarray}
{\cal A}^{\rm NP}\sim d^{2}_{(h_{W^-}-h_{W^+}),(h_{q}-h_{\bar{q}})},\quad {\cal A}^{\rm SM}\sim d^{1}_{(h_{W^-}-h_{W^+}),(h_{q}-h_{\bar{q}})}.\label{eq:dmatrix}
\end{eqnarray}
By the orthogonality of the $d$-matrix, the interference amplitude vanishes after integrating over the solid angle.

Another interesting observation is that, for ${\cal O}_{16}$ and ${\cal O}_{17}$, the integrated interference amplitude scales as $S$, rather than $S^2$ before the integration. This can also be understood with the help of partial wave decomposition. From the J-basis analysis, both of the operators generate scattering amplitude for $q\bar{q}\to G^+ G^-$ with angular momentum $J=2$, where $G^{\pm}$ is the Goldstone corresponds to the longitudinal mode of $W^\pm$. 
However, if one expands the $(0,0)$ polarization component of the SM amplitude in a series of powers of $S$, one can find that the coefficient for $S^0$ term is proportional to $\cos\theta$ which corresponds to a partial wave of $J=1$, the $J=2$ partial wave is contained in the term of expansion order of $S^{-1}$. 
Such a property is more transparent in the Feynman gauge, where the $(0,0)$ polarization is generated by the s-channel diagram with the virtual gauge boson decaying to a pair of charged Goldstone bosons, which contributes only to the $J=1$ partial wave. The SM t-channel diagram which contains the partial wave of $J=2$ involves only the transverse polarization final states\footnote{The fermion to Goldstone couplings are suppressed by Yukawa couplings.}, and to convert it into the  $(0,0)$ polarization state two insertions of Goldstone boson-gauge boson vertices are needed, which decrease the dependence on the scattering energy by $1/S$.

\begin{table}[h]
\begin{center}
\renewcommand{\arraystretch}{1.5}
\begin{tabular}{ |c|c|c| } 
 \hline
 Operator & $ 2\operatorname{Re}({\cal A}^{\rm SM}{\cal A}^{\rm NP *})$ & $2\int  d\Omega \operatorname{Re}({\cal A}^{\rm SM}{\cal A}^{\rm NP *})$  \\ 
 \hline
 ${\cal O}_8$ & $d\bar{d}:\  b_8 S+c_8
 $ & 0\\ 
 \hline
 ${\cal O}_9$ & $\bar{u}u:\  b_9 S+c_9$  & 0\\ 
 \hline
 \multirow{1}{*}{${\cal O}_{10}$} &  \makecell{$u\bar{u}/d\bar{d}:\   a_{10}\cdot S^2+b_{10}\cdot S + c_{10}$}& {$u\bar{u}/d\bar{d}:\ \ov{ a}_{10}\cdot S^2+\ov{b}_{10}\cdot S + \ov{c}_{10}$}\\
 \hline
  ${\cal O}_{11}$ &  \makecell{0} &0\\ 
 \hline
 \multirow{2}{*}{${\cal O}_{12}$} &  \makecell{$u\bar{u}:\  a^{u}_{12}S^2+b^{u}_{12}S+c^{u}_{12}$} & \makecell{$u\bar{u}:\ \ov{ a}^{u}_{12}S^2+\ov{b}^{u}_{12}S+\ov{c}^{u}_{12}+\ov{D}^{u}_{12}\log{S}$}\\ 
 \cline{2-3}
  &\makecell{$d\bar{d}:\  a^{d}_{12}S^2+b^{d}_{12}S+c^{d}_{12}$} & \makecell{$d\bar{d}:\ \ov{ a}^{d}_{12}S^2+\ov{b}^{d}_{12}S+\ov{c}^{d}_{12}+\ov{D}^{d}_{12}\log{S}$}\\
 \hline
  \multirow{2}{*}{${\cal O}_{13}$} &  \makecell{$u\bar{u}:\  a^{u}_{13}S^2+b^{u}_{13}S+c^{u}_{13}$} & \makecell{$u\bar{u}:\ \ov{ a}^{u}_{13}S^2+\ov{b}^{u}_{13}S+\ov{c}^{u}_{13}+\ov{D}^{u}_{13}\log{S}$}\\ 
  \cline{2-3}
  &  \makecell{$d\bar{d}:\  a^{d}_{13}S^2+b^{d}_{13}S+c^{d}_{13}$} & \makecell{$d\bar{d}:\ \ov{ a}^{d}_{13}S^2+\ov{b}^{d}_{13}S+\ov{c}^{d}_{13}+\ov{D}^{d}_{13}\log{S}$}\\ 
 \hline
 ${\cal O}_{14}$ & $u\bar{u}:\   a_{14}S^2+b_{14} S+c_{14}
 $ & 0\\ 
 \hline
  ${\cal O}_{15}$ & $d\bar{d}:\   a_{15}S^2+b_{15} S+c_{15} 
 $ & 0\\ 
 \hline
 \multirow{2}{*}{${\cal O}_{16}$} &  \makecell{$u\bar{u}:\  a^{u}_{16}S^2+b^{u}_{16}S+c^{u}_{16}$} & \makecell{$u\bar{u}:\ \ov{b}^{u}_{16}S+\ov{c}^{u}_{16}+\ov{D}^{u}_{16}\log{S}$}\\ 
  \cline{2-3}
  &  \makecell{$d\bar{d}:\  a^{d}_{16}S^2+b^{d}_{16}S+c^{d}_{16}$} & \makecell{$d\bar{d}:\ \ov{b}^{d}_{16}S+\ov{c}^{d}_{16}+\ov{D}^{d}_{16}\log{S}$}\\ 
 \hline
  \multirow{2}{*}{${\cal O}_{17}$} &  \makecell{$u\bar{u}:\  a^{u}_{17}S^2+b^{u}_{17}S+c^{u}_{17}$} & \makecell{$u\bar{u}:\ \ov{b}^{u}_{17}S+\ov{c}^{u}_{17}+\ov{D}^{u}_{17}\log{S}$}\\ 
  \cline{2-3}
  &  \makecell{$d\bar{d}:\  a^{d}_{17}S^2+b^{d}_{17}S+c^{d}_{17}$} & \makecell{$d\bar{d}:\ \ov{b}^{d}_{17}S+\ov{c}^{d}_{17}+\ov{D}^{d}_{17}\log{S}$}\\ 
 \hline
\end{tabular}\caption{Scaling of $q\bar{q}\to WW$ interference amplitude after summing and averaging over spins and helicities.}\label{tab:scalingWW}
\end{center}
\end{table}

\begin{table}[htb]
\begin{center}
\renewcommand{\arraystretch}{1.5}
\begin{tabular}{ |c|c|c| } 
 \hline
 $(h_{W^+}, h_{W^-})$ &${\cal A}_{h_i}^{\rm 8}/\frac{C_8}{\Lambda^4}$  & ${\cal A}_{h_i}^{\rm SM}$  \\ 
 \hline
 $-,+$ & $  S \sin ^2\left(\frac{\theta
   }{2}\right) \sin (\theta )
    \left(S-2
   M_W^2\right)\delta _{a b}$
  & 0\\  
 \hline
  $-,-$ & $  S \sin (\theta ) \cos (\theta ) M_W^2 \delta _{a b}$
  & $\frac{4 \pi  \alpha \sin (\theta ) M_Z^2 \delta _{a b} \sqrt{1-\frac{4 M_W^2}{S}}}{3 \left(S-M_Z^2\right)}$\\ 
 \hline 
  $-,0$ & $\frac{  S^{3/2} \sin ^2\left(\frac{\theta }{2}\right) (2 \cos (\theta )+1) M_W^2 \delta _{a b}}{\sqrt{2} M_W}$
  & $\frac{4 \pi  \alpha  \sin ^2\left(\frac{\theta }{2}\right) M_Z^2 \delta _{a b} \sqrt{2 S-8 M_W^2}}{3 S M_W-3 M_W M_Z^2}$\\ 
 \hline 
  $+,+$ & $  S \sin (\theta ) \cos (\theta ) M_W^2 \delta _{a b}$
  & $\frac{4 \pi  \alpha  \sin (\theta ) M_Z^2 \delta _{a b} \sqrt{1-\frac{4 M_W^2}{S}}}{3 \left(S-M_Z^2\right)}$\\ 
 \hline 
  $+,-$ & $-2   S \sin \left(\frac{\theta }{2}\right) \cos ^3\left(\frac{\theta }{2}\right) \delta _{a b} \left(S-2 M_W^2\right)$
  & 0\\ 
 \hline
   $+,0$ & $\frac{  S^{3/2} \cos ^2\left(\frac{\theta }{2}\right) (2 \cos (\theta )-1) M_W^2 \delta _{a b}}{\sqrt{2} M_W}$
  & $\frac{4 \pi  \alpha  \cos ^2\left(\frac{\theta }{2}\right) M_Z^2 \delta _{a b} \sqrt{2 S-8 M_W^2}}{3 S M_W-3 M_W M_Z^2}$\\ 
 \hline
   $0,+$ & $-\frac{  S^{3/2} \sin ^2\left(\frac{\theta }{2}\right) (2 \cos (\theta )+1) M_W^2 \delta _{a b}}{\sqrt{2} M_W}$
  & $\frac{4 \sqrt{2} \pi  \alpha  \sin ^2\left(\frac{\theta }{2}\right) M_Z^2 \delta _{a b} \sqrt{S-4 M_W^2}}{3 M_W M_Z^2-3 S M_W}$\\ 
 \hline
   $0,-$ & $\frac{  S^{3/2} \cos ^2\left(\frac{\theta }{2}\right) (1-2 \cos (\theta )) M_W^2 \delta _{a b}}{\sqrt{2} M_W}$
  & $\frac{4 \sqrt{2} \pi  \alpha  \cos ^2\left(\frac{\theta }{2}\right) M_Z^2 \delta _{a b} \sqrt{S-4 M_W^2}}{3 M_W M_Z^2-3 S M_W}$\\ 
 \hline
    $0,0$ & $-   S  \sin2\theta M_W^2 \delta _{a b}$
  & $\frac{2 \pi  \alpha  \sin (\theta ) M_Z^2 \delta _{a b} \left(2 M_W^2+S\right) \sqrt{1-\frac{4 M_W^2}{S}}}{3 M_W^2 \left(M_Z^2-S\right)}$\\ 
 \hline
\end{tabular}\caption{Helicity amplitudes for $d\bar{d}\to WW$ for $h_{d}=1$ and $h_{\bar{d}}=-1$, where ${\cal A}_{h_i}^{\rm 8}$ is generated by ${\cal O}_8$.}\label{tab:O8hel}
\end{center}
\end{table}

\begin{table}[htb]
\begin{center}
\renewcommand{\arraystretch}{1.5}
\begin{tabular}{ |c|c| } 
 \hline
 $(h_{W^+}, h_{W^-})$ &${\cal A}_{h_i}^{\rm 15}/\frac{C_{15}}{\Lambda^4}$    \\ 
 \hline
 $-,+$ & $  S \sin ^2\left(\frac{\theta }{2}\right) \sin (\theta ) M_W^2 \delta _{a b}$
  \\  
 \hline
  $-,-$ & $-  S \sin ^2\left(\frac{\theta }{2}\right) \sin (\theta ) M_W^2 \delta _{a b}$
  \\ 
 \hline 
  $-,0$ & $\frac{  S^{3/2} \sin ^2\left(\frac{\theta }{2}\right) \cos (\theta ) M_W \delta _{a b}}{\sqrt{2}}$
 \\ 
 \hline 
  $+,+$ & $  S \sin (\theta ) \cos ^2\left(\frac{\theta }{2}\right) M_W^2 \delta _{a b}$
 \\ 
 \hline 
  $+,-$ & $-  S \sin (\theta ) \cos ^2\left(\frac{\theta }{2}\right) M_W^2 \delta _{a b}$
  \\ 
 \hline
   $+,0$ & $\frac{  S^{3/2} \cos ^2\left(\frac{\theta }{2}\right) \cos (\theta ) M_W \delta _{a b}}{\sqrt{2}}$
  \\ 
 \hline
   $0,+$ & $-\frac{  S^{3/2} \sin ^2(\theta ) M_W \delta _{a b}}{2 \sqrt{2}}$
  \\ 
 \hline
   $0,-$ & $\frac{  S^{3/2} \sin ^2(\theta ) M_W \delta _{a b}}{2 \sqrt{2}}$
  \\ 
 \hline
    $0,0$ & $-\frac{1}{8}   S^2 \sin (2 \theta ) \delta _{a b}$
  \\ 
 \hline
\end{tabular}\caption{Helicity amplitudes for $d\bar{d}\to WW$ for $h_{d}=1$ and $h_{\bar{d}}=-1$, where ${\cal A}_{h_i}^{\rm 15}$ is generated by ${\cal O}_{15}$.}\label{tab:O15hel}
\end{center}
\end{table}

\begin{figure}[h]
\centering
\includegraphics[width=0.49\textwidth]{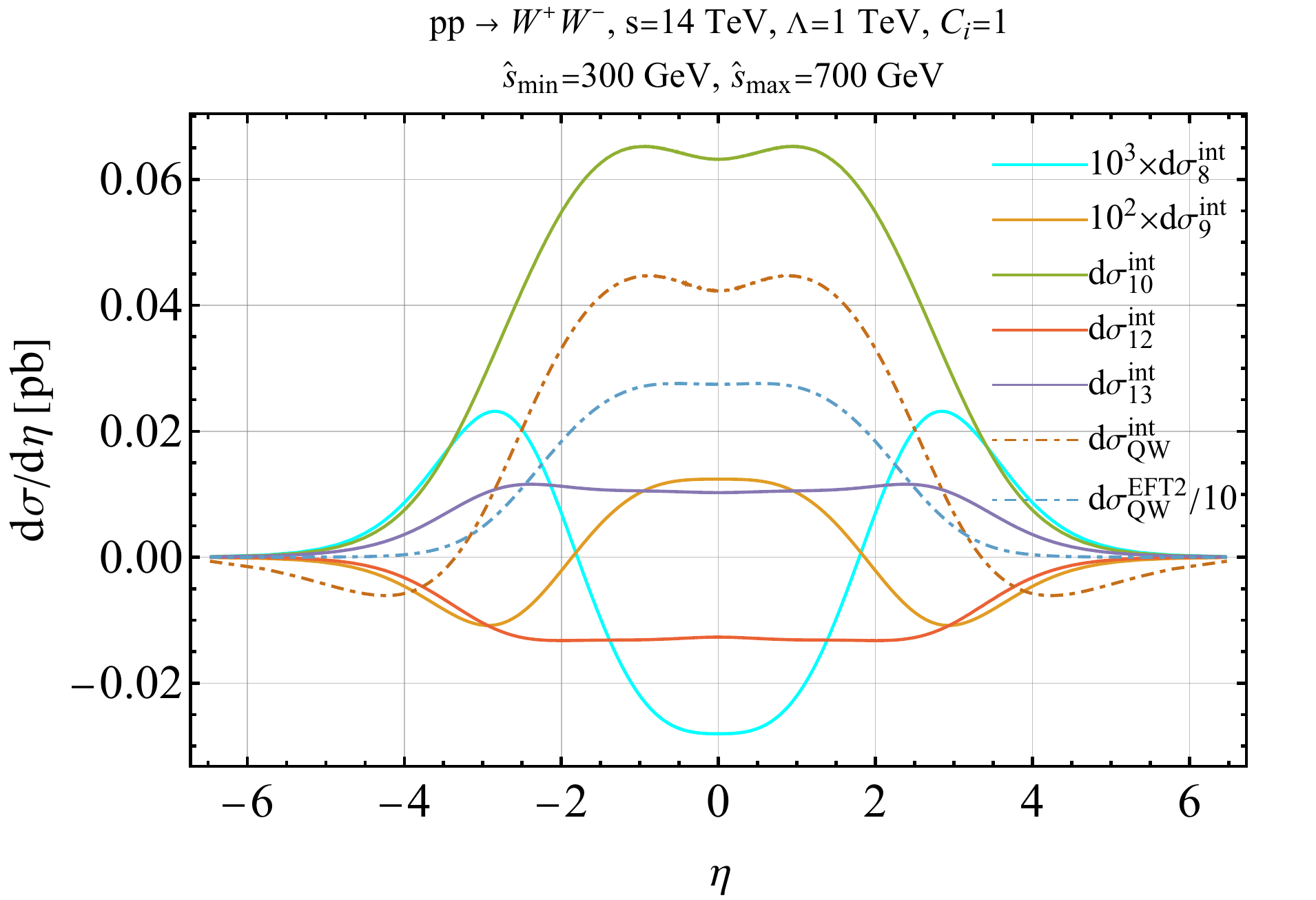}
\includegraphics[width=0.49\textwidth]{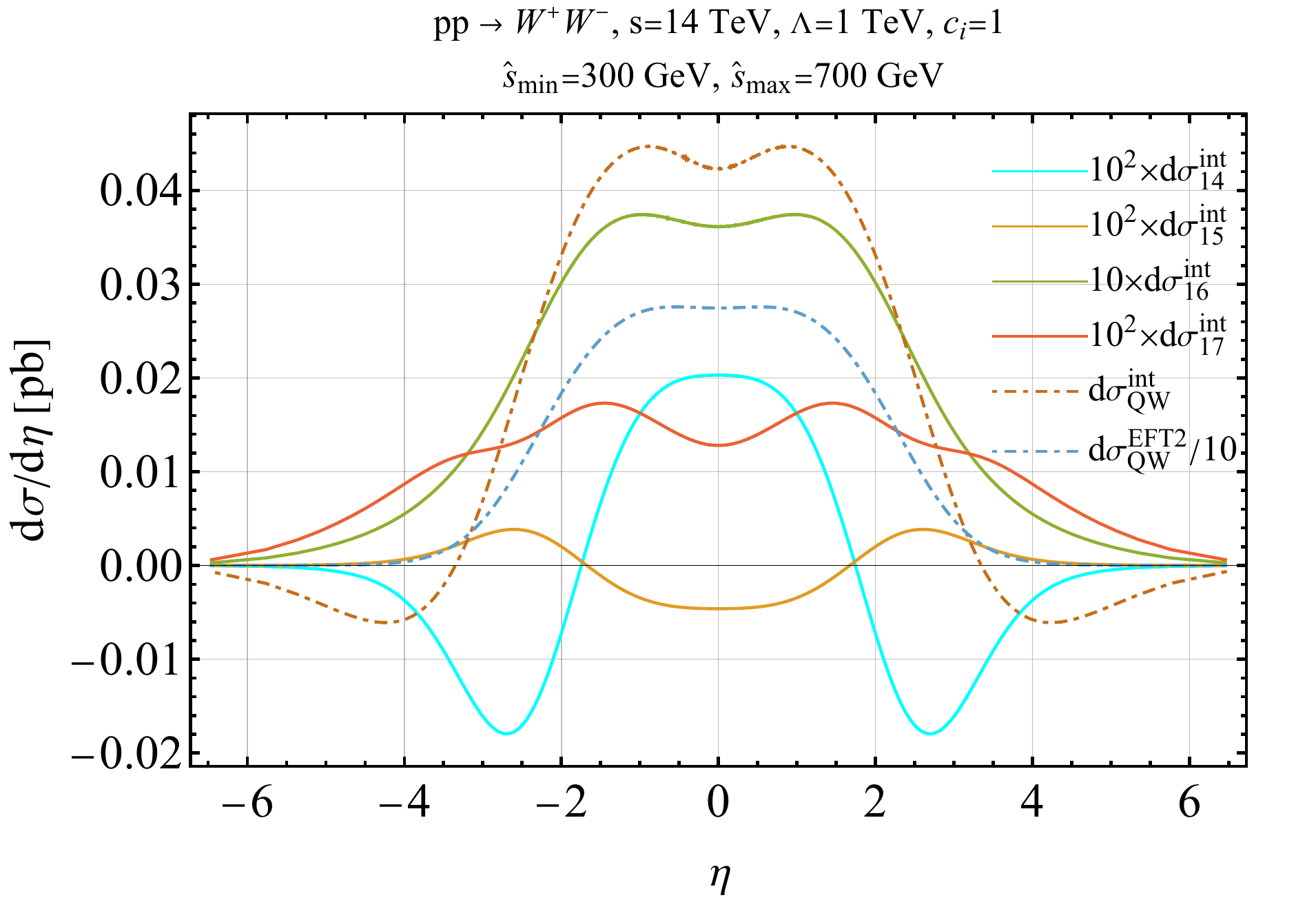}
\caption{Differential cross-sections for dimension-8 and dimension-6 operators. The solid lines are for interference cross-sections for the dim-8 operators, while dash-dotted brown and blue lines are for the ${\rm BSM}\times {\rm SM}$ interference and the ${\rm BSM}^2$ generated by the dimension-6 ${\cal O}_{W}$ respectively. The results for $O_{8,9}$ operators are scaled by a factor of $100$. The result for ${\rm BSM}^2$ is rescaled by a factor of $0.1$, and taking into account the loop-generated nature of $O_W$, in reality it could be further suppressed compared to the $\text{dim-8}\times {\rm SM}$ contributions due to an additional loop factor.}\label{fig:dsigWW}
\end{figure}

In figure~\ref{fig:dsigWW}, we plot the hadronic differential cross-section with respect to the scattering angle in the diboson center of mass frame, $\theta$,  at the 14 TeV LHC. We set the new physics scale at $\Lambda=1$ TeV as an example and set all the dimensionless Wilson coefficients equal to 1. 
We also add as a comparison the contributions from dim-6 interference and dim-6 square differential cross-section for the  operator ${\cal O}_W$. 
In principle, the $O_W$ operator can only be generated through loop-level matching. As a result, the dimensionless Wilson coefficient tends to receive a $1/(16\pi^2)$ suppression. Nevertheless, we still set $c_i$ to unity to compare the relative contribution from different operators, the concrete scale of each Wilson coefficient depends on the physical assumption of the UV model, which is beyond the scope of our study in this paper.
In the plot, we conservatively rescale the dim-6 square contribution by a factor of 0.1 to show its shape. In fact, if the UV theory responsible for the ${\cal O}_W$ is weakly coupled, then the dim-6 square contribution will at least be suppressed by a loop factor $1/(16\pi^2)\sim 0.01$ compared to the dim-8 and SM interference by naive dimension analysis.

The dominance of the dim-6 square contribution is coming from the enhancement in the amplitude scaling as $S^2$ in the high energy limit. 
In the left panel, the three solid lines in green, red, and purple represent the differential interference cross-sections for operators ${\cal O}_{10}$, ${\cal O}_{12}$ and ${\cal O}_{13}$ respectively, and they all have different angular distributions comparing with each other and with the dim-6 originated contributions. 
Therefore in principle, one can measure the angular distribution of the vector boson in the diboson production to differentiate effects from dim-8 SMEFT operators with dim-6 ones. 
On the other hand, the distributions for ${\cal O}_{12}$ and ${\cal O}_{13}$ look similar except for a global sign, thus presenting a difficulty in discerning the two operators.

\begin{figure}[h]
\centering
\includegraphics[width=0.49\textwidth]{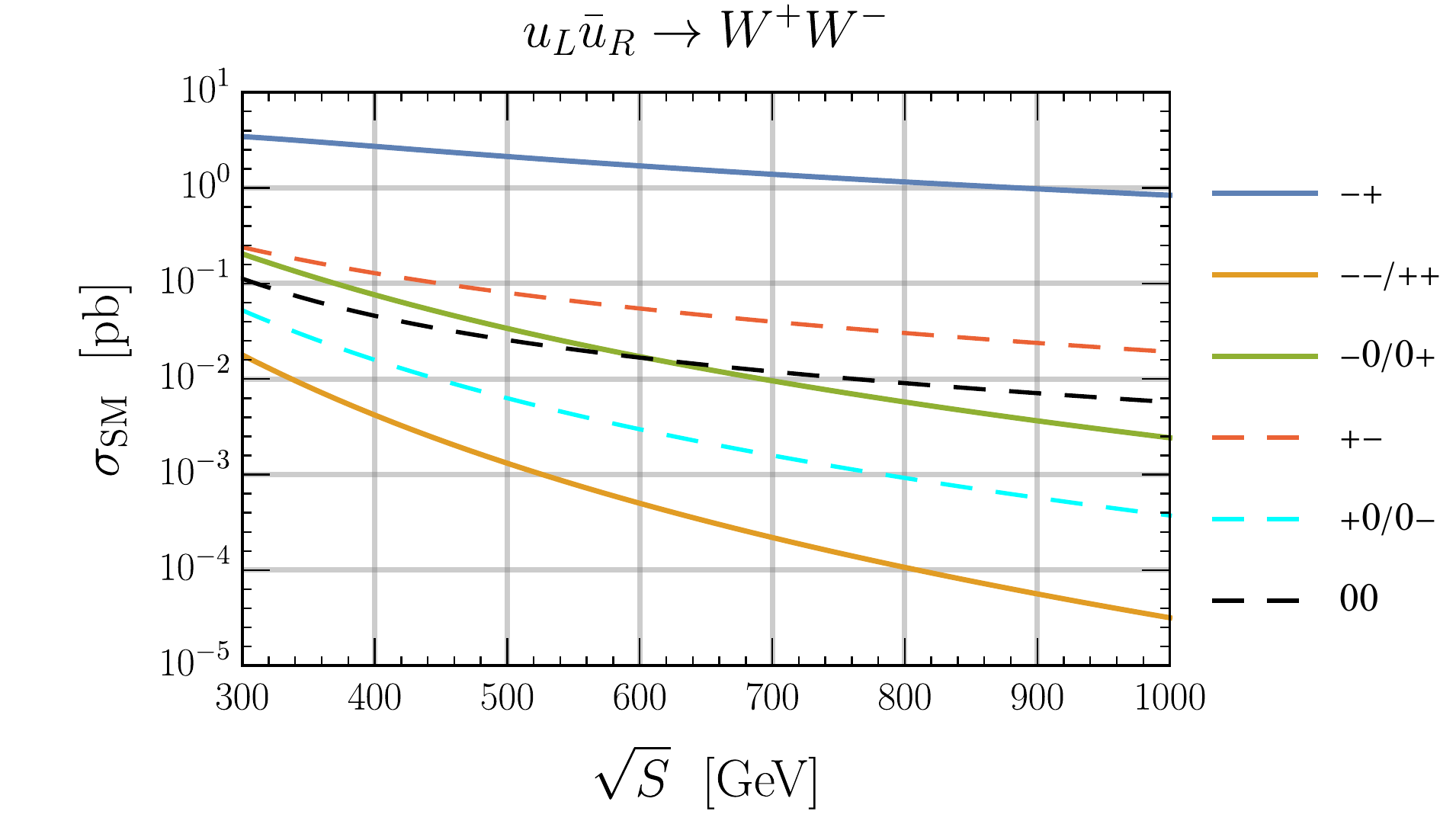}
\includegraphics[width=0.49\textwidth]{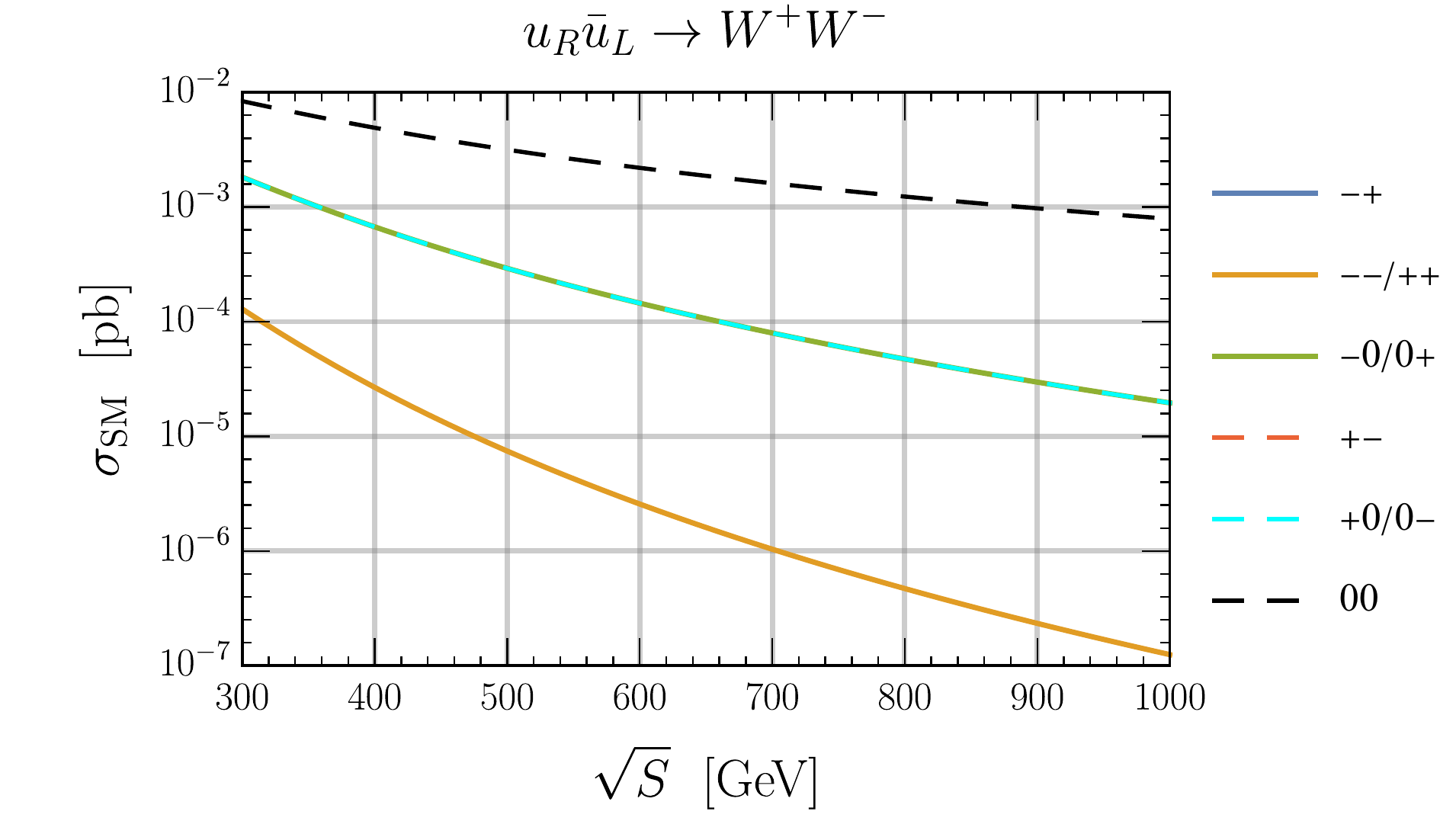}
\includegraphics[width=0.49\textwidth]{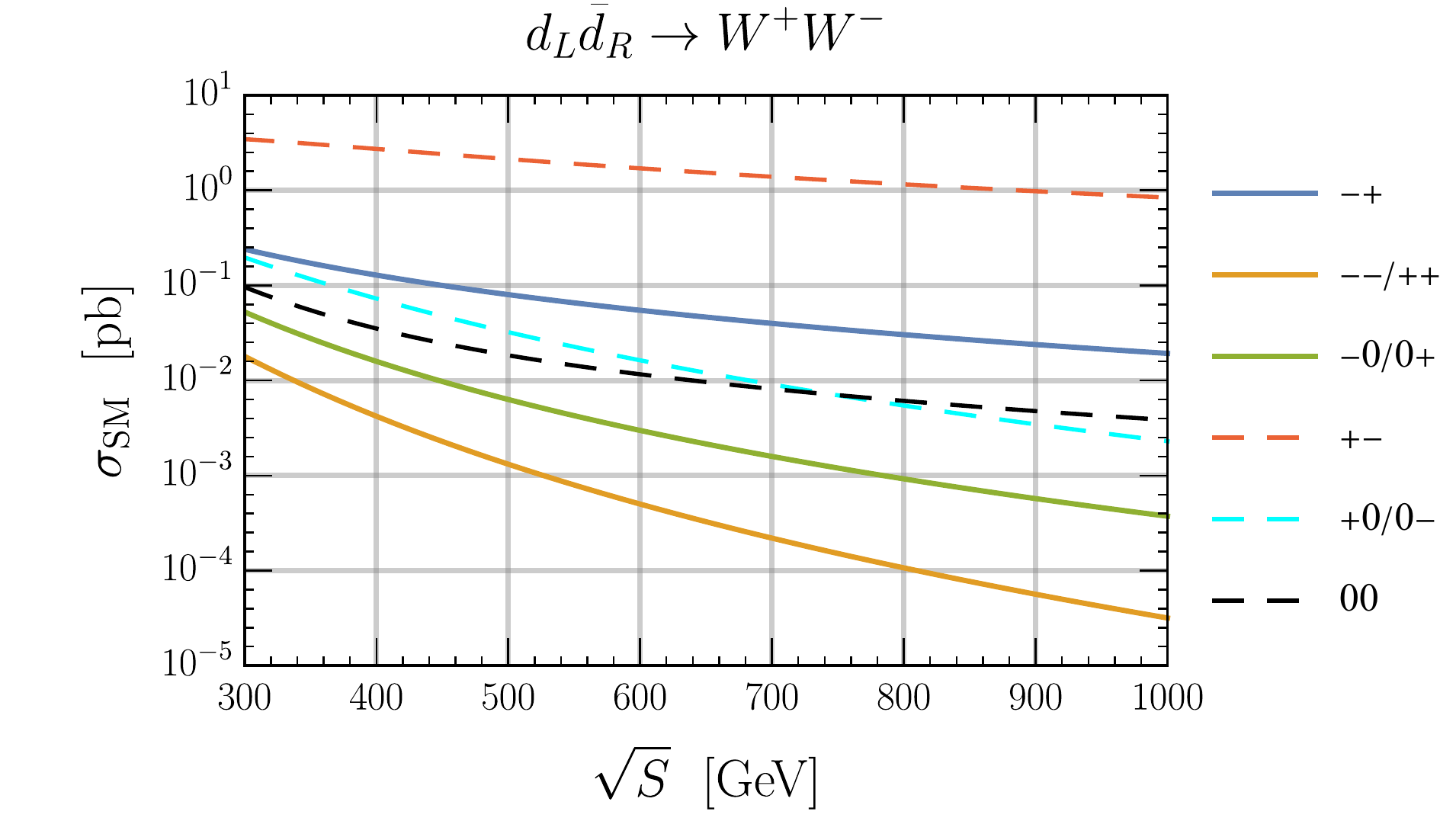}
\includegraphics[width=0.49\textwidth]{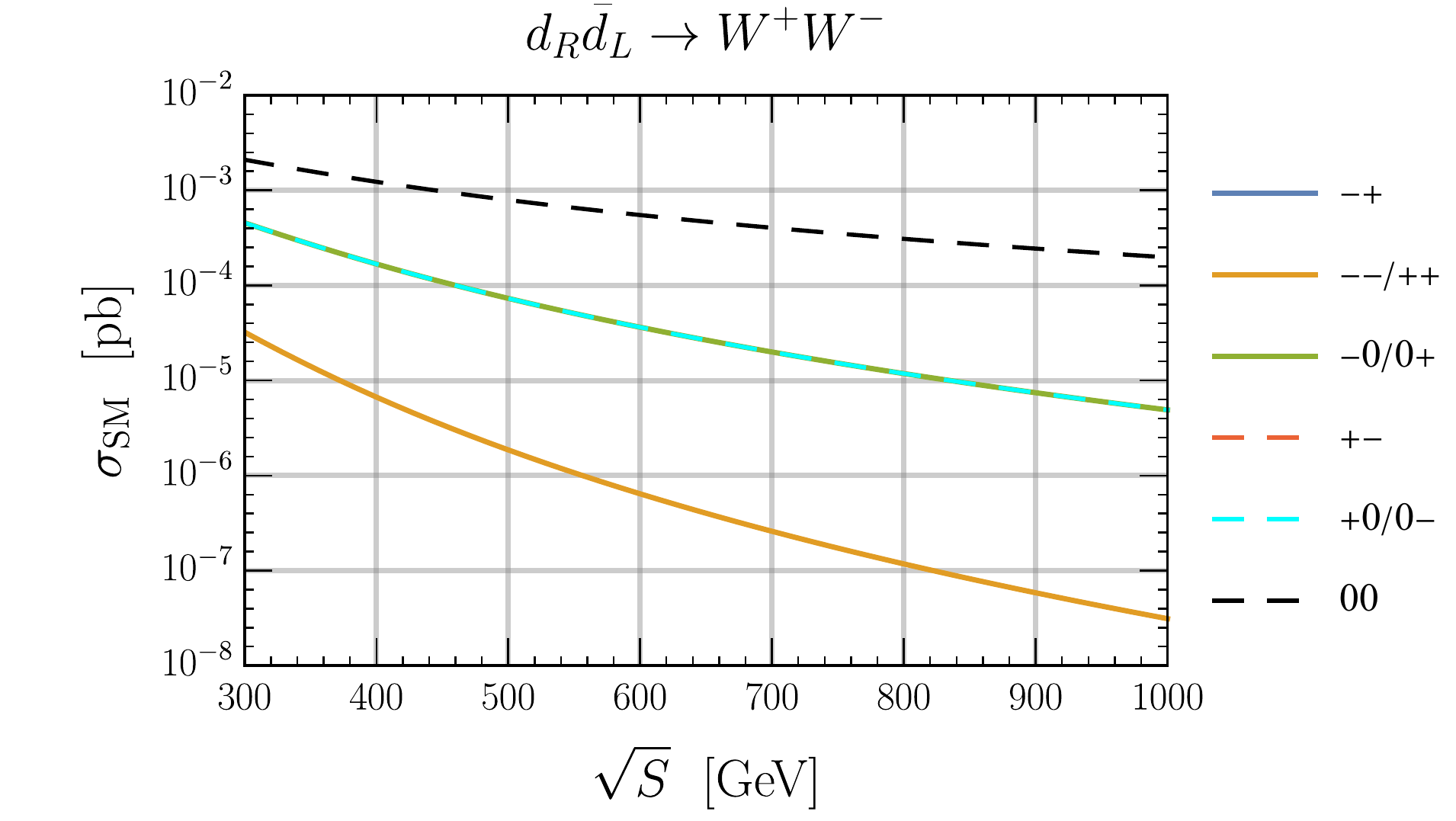}
\caption{SM parton level cross-section as a function of the center of mass frame energy for different helicity of W bosons. The upper panel is for the $u$ quark-initiated process, and the bottom panel is for the $d$ quark-initiated process. Different colors represent different helicities of $W^+$ and $W^-$, respectively. In the right panel, the cross-section for helicity states $-+$ and $+-$ for right-handed quark-initiated processes are zero and thus are not visible on the log-scale plots.}\label{fig:smhelsep}
\end{figure}

In the right panel in the figure~\ref{fig:dsigWW}, we plot the similar differential cross-section for the operators that contain the Higgs field --- ${\cal O}_{14-17}$. One can find that the scales of the interference cross-sections are generally smaller than those generated by operators without Higgs field by one to two orders of magnitude for $O_{16,17}$ compared to $O_{10- 13}$.
The reason is that only the $(0,0)$ polarized final state gets enhanced by $S^2$ in the high energy regime for ${\cal O}_{16,17}$, while the SM amplitude of two longitudinal polarized $W$ bosons for the left-handed initiated process is generally smaller than those of the $(+,-)$ or $(-,+)$ polarized final states as one can see from the left panel in figure~\ref{fig:smhelsep}, where we plot the SM parton level cross-sections of each helicity configurations as functions of the center of mass energy.

\begin{figure}[h]
\centering
\includegraphics[width=0.6\textwidth]{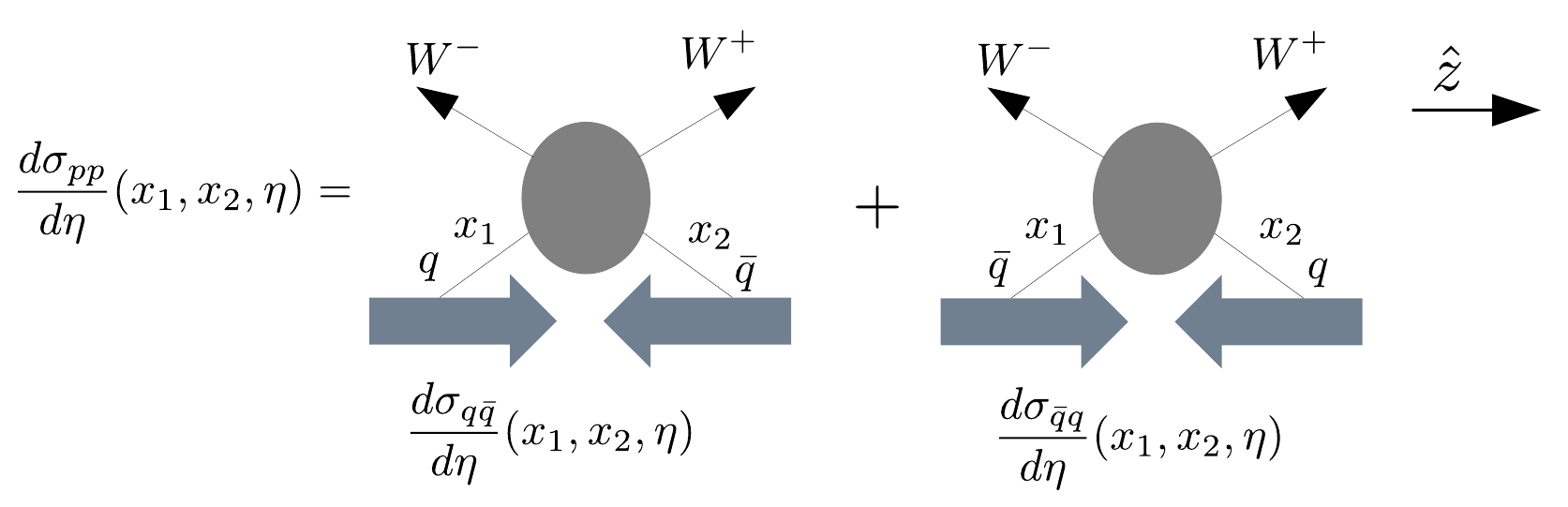}
\caption{The pictorial illustration for the relation between the total differential cross-section and the partonic differential cross-sections}\label{fig:dsig_eta}
\end{figure}

\begin{figure}[h]
\centering
\includegraphics[width=0.95\textwidth]{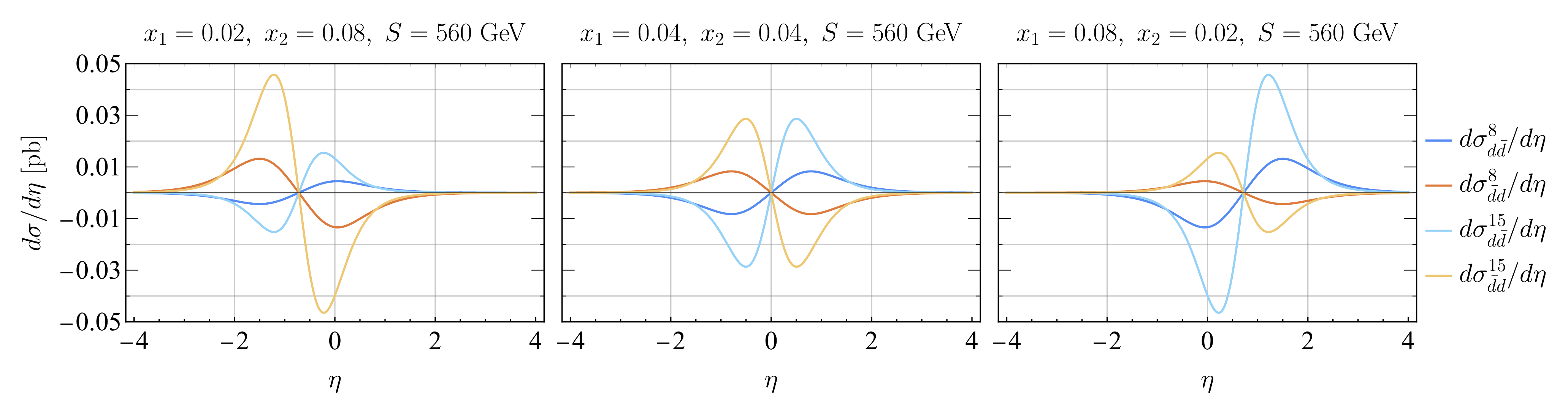}
\caption{Comparison of the differential cross-section for $q\bar{q}$ and $\bar{q}q$ channel for $O_8$ and $O_{15}$, as one can see from the middle plot, the total differential cross-section vanished in the COM frame of partons as $q\bar{q}$ and $\bar{q}q$ contributions cancel exactly.}\label{fig:d8eta}
\end{figure}

On the contrary, the corresponding differential interference cross-section for ${\cal O}_{14,15}$ and ${\cal O}_{8,9}$ are at a similar order even though for the right-handed quark-initiated process, the amplitude of the $(0,0)$ polarized final state is generally larger than those of the final states with transverse polarization as one can see from the right panel in figure~\ref{fig:smhelsep}.
This results from the fact that the $q\bar{q}$ and $\bar{q}q$ processes get strong cancellations for these four operators, such that a simple order of magnitude comparison for individual amplitude is not valid anymore. 
The relation  between the total pp differential cross-section and the separated partonic cross-sections are illustrated in figure~\ref{fig:dsig_eta}. The definition of the differential cross-section with respect to the pseudo-rapidity of $W^+$ in the COM frame of the pp system is detailed in appendix~\ref{app:dsignoncom}. 
To illustrate the cancellation between $q\bar{q}$ and $\bar{q}q$ channels, we plot in figure~\ref{fig:d8eta} the interference differential cross-sections $d\sigma_{q\bar{q}}/d\eta(x_1,x_2,\eta)$ and  $d\sigma_{\bar{q}q}/d\eta(x_1,x_2,\eta)$ for $q\bar{q}$ and $\bar{q}q$ processes respectively, where  $x_1$ and $x_2$ indicate the momentum fractions for $q$   and $\bar{q}$  for the $q\bar{q}$  process and vice verse for the $\bar{q}q$ process, and $\eta$ is the pseudo-rapidity of the $W^+$ in the lab frame. 
From figure~\ref{fig:d8eta}, one can find that in the center of mass frame of the parton system, the cross-section from $q\bar{q}$ and $\bar{q}q$ channels cancel each other exactly, and as one boosts along the beam direction, such cancellation is restored by the difference between the probabilities of finding a $d$-quark and $\bar{d}$-quark in the proton as shown in the left and the right plots in the figure.

\begin{figure}
\centering
         \includegraphics[width=0.3\textwidth]{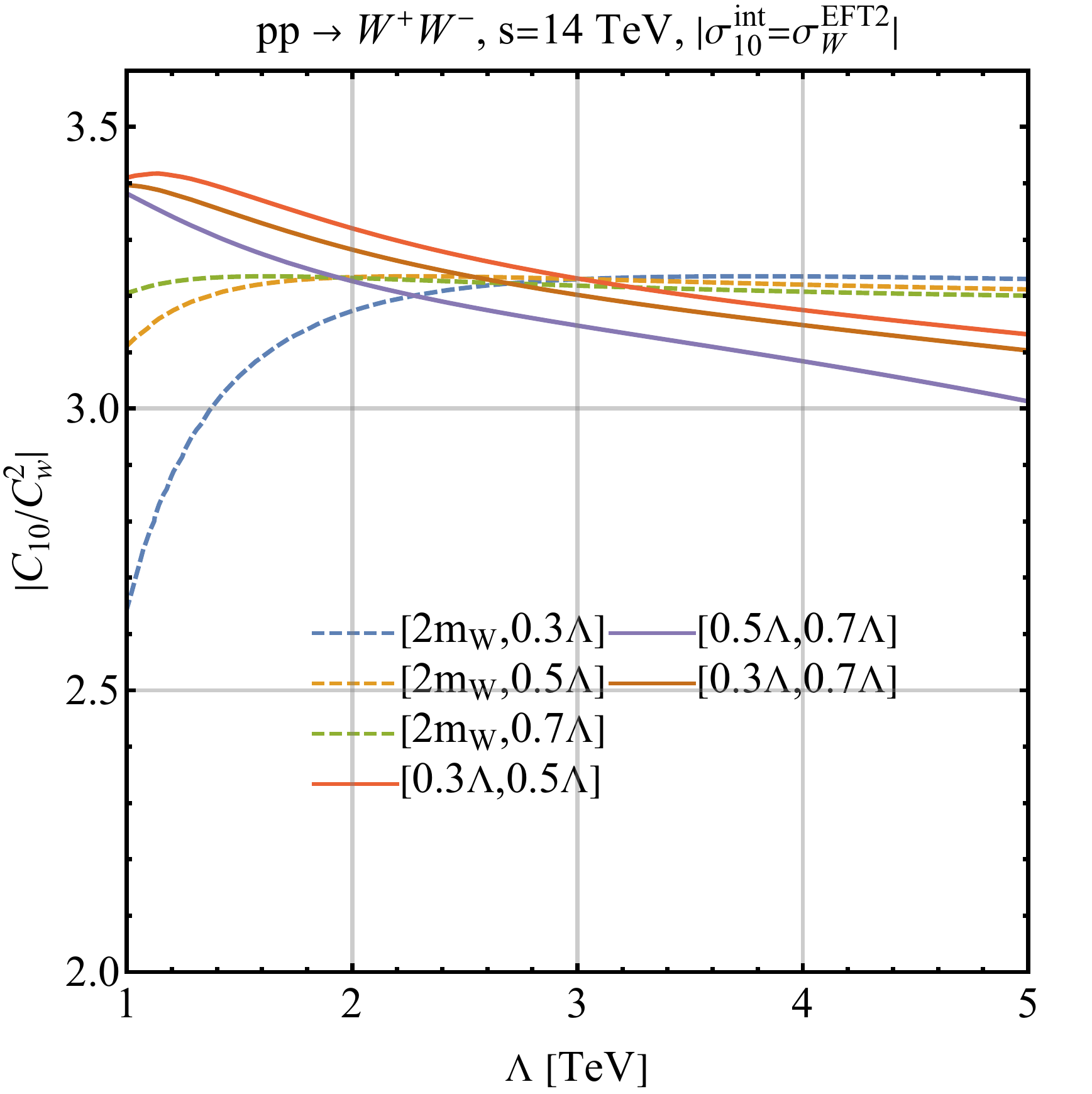}
         \includegraphics[width=0.3\textwidth]{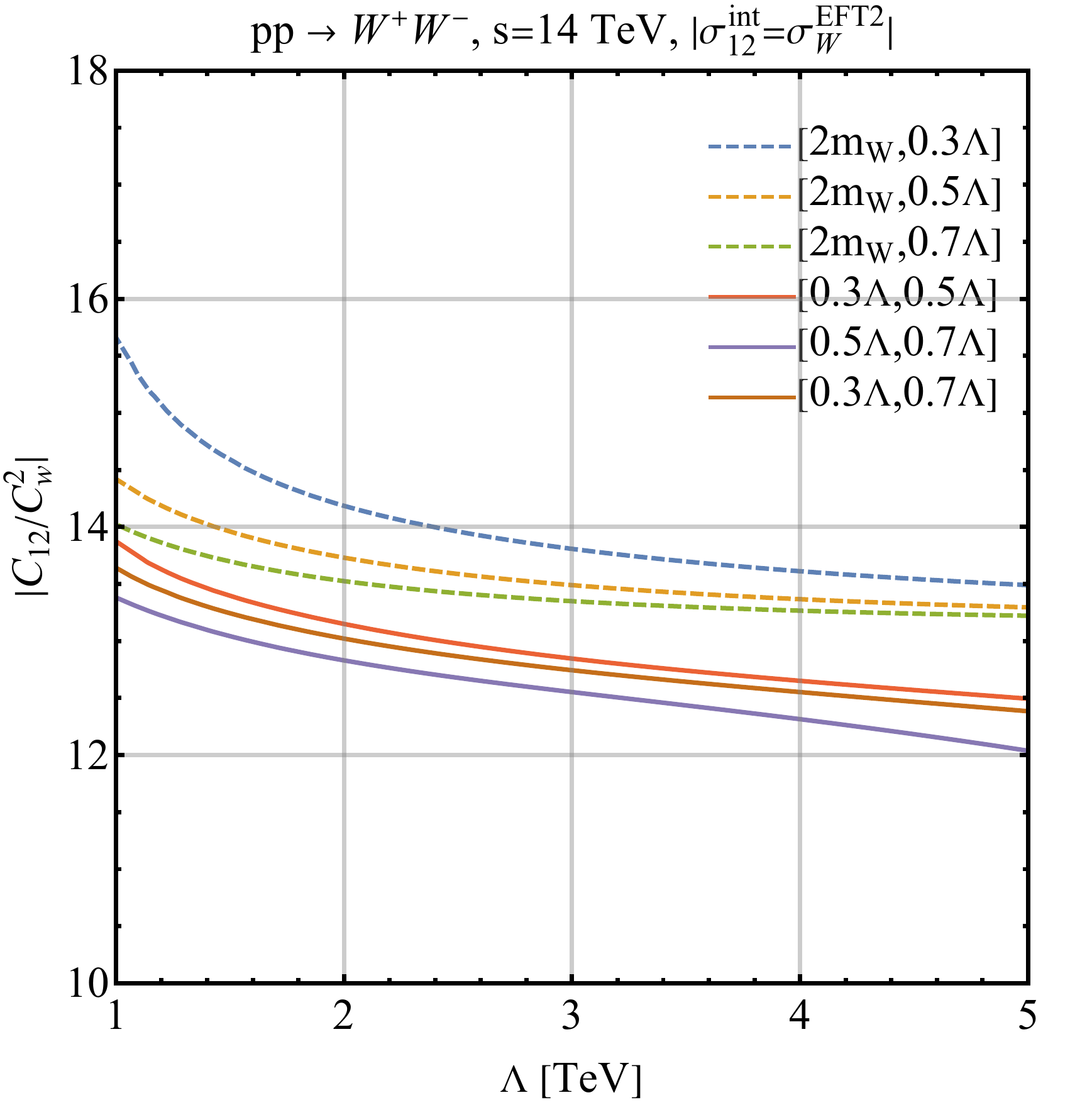}
         \includegraphics[width=0.3\textwidth]{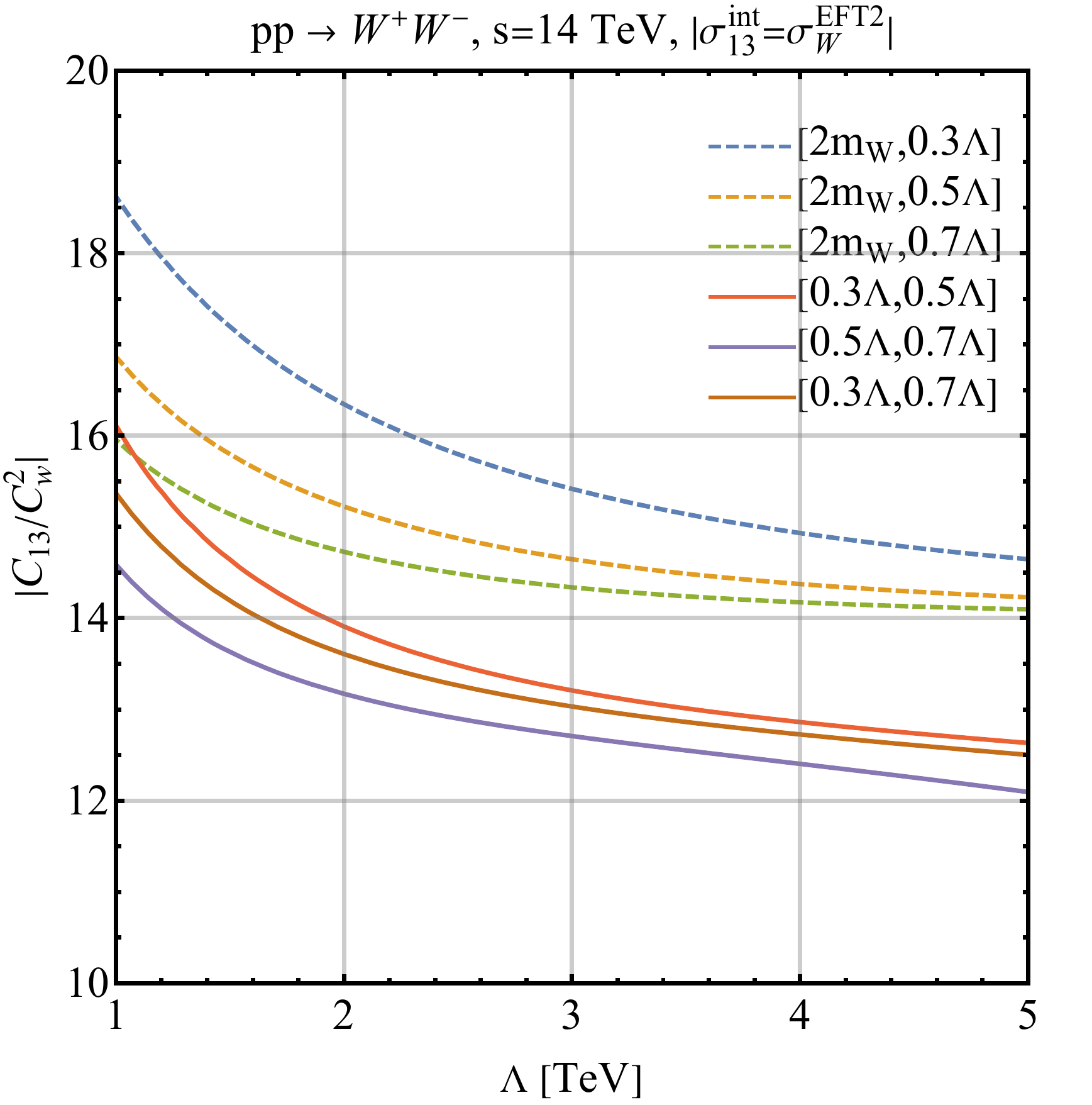}
         \includegraphics[width=0.3\textwidth]{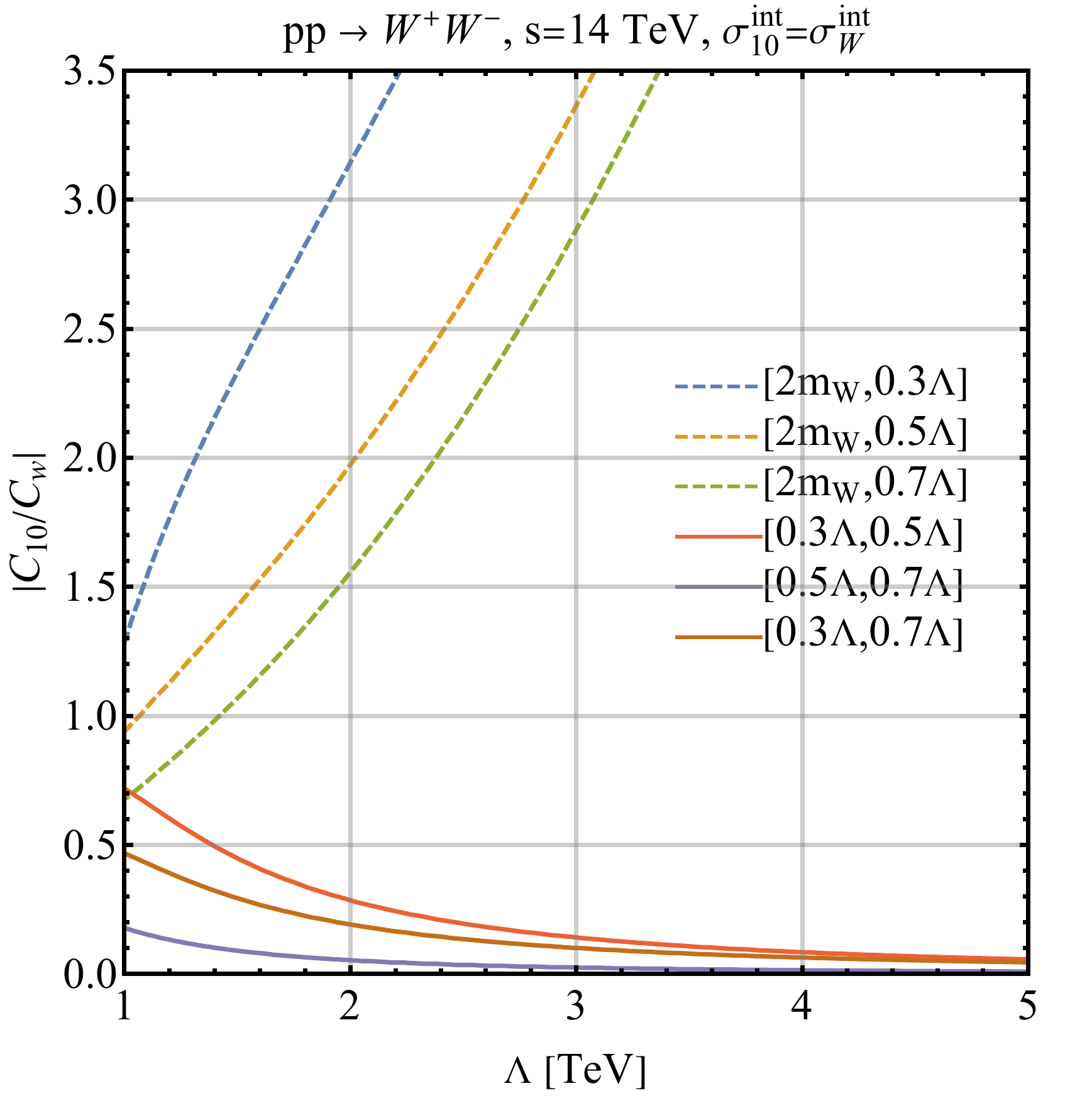}
         \includegraphics[width=0.3\textwidth]{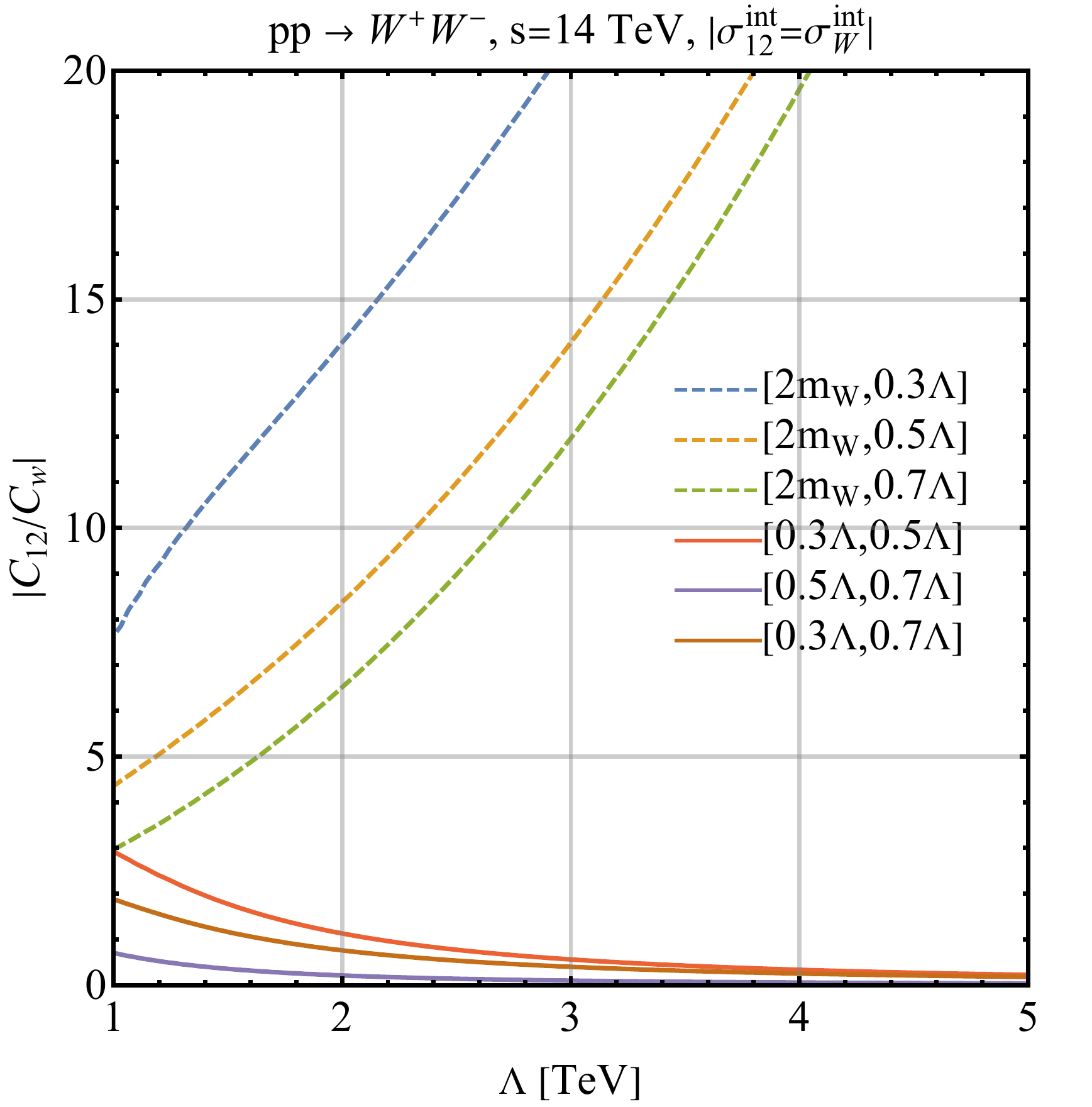}
         \includegraphics[width=0.3\textwidth]{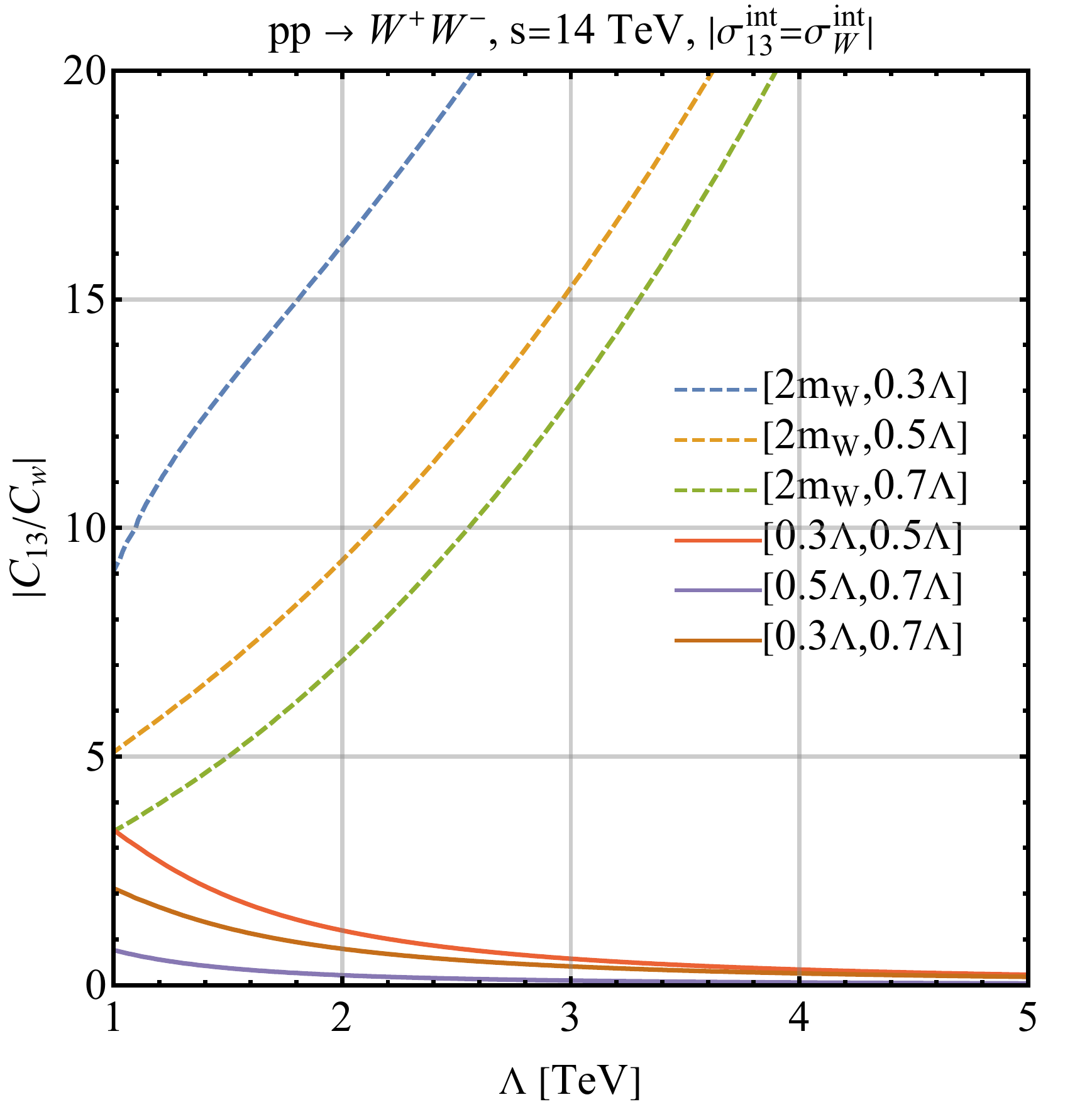}
\caption{Comparisons of the interference cross-sections of dim-8 operator ${\cal O}_{10}$, ${\cal O}_{12}$ and ${\cal O}_{13}$ respectively from left to right with respect to the cross-section from the dim-6 amplitude square (top) and dim-6 interference (bottom). Different colors represent different cuts on the diboson invariant mass, the lower bonds for the dashed lines are fixed at the diboson production threshold.  The dim-8 interference cross-section and dim-6 square cross-section (top) or dim-6 interference cross-section (bottom) have equal magnitude for each point on the lines.}\label{fig:WWEFT2}
\end{figure}

In figure~\ref{fig:WWEFT2}, we demonstrate the relative scale of the diboson cross-section from the dim-8-SM interference term to those from the dim-6 square term (dim-6-SM interference term) in the top (bottom) row. 
In each plot, each curve is obtained by equating the magnitudes of the two cross-sections in the comparison. Different colored lines represent  different cuts on the diboson invariant mass $M_{WW}$, and the dashed lines all have a fixed lower bond as the kinematic threshold for the diboson production while varying the upper bounds with the NP scale $\Lambda$.
The solid lines, however, also vary the lower bounds with respect to the NP scale $\Lambda$. From the plot, one can deduce that the effect from ${\cal O}_{10}$ is larger than those from ${\cal O}_{12}$ and ${\cal O}_{13}$ with equal dimensionless Wilson coefficients. 
This can be seen from the fact that one needs a relatively smaller ratio of dimensionless Wilson coefficients $|C_{10}/C^2_{W}|\sim 3$ for ${\cal O}_{10}$ to achieve the equal magnitude of the interference cross-section comparing to the dim-6 square counterpart, 
while for ${\cal O}_{12}$ or ${\cal O}_{13}$ a ratio of order ${\cal O}(10)$ is needed.  
This fact indicates that when people include the dim-6 square effects in their global analysis in the SMEFT framework, they should also take into account at least the effect from ${\cal O}_{10}$. Furthermore, as discussed before, if one assumes the operator ${\cal O}_{W}$ is generated by a weakly coupled UV theory, then $O_i$ with $|C_{i}/C^2_{W}|\sim 100$ should also be considered to be important.

From the plots in the first row in figure~\ref{fig:WWEFT2}, one can also observe that the dashed lines become flat at a high NP scale, which can be understood as follows. Both  $\sigma_i^{\rm int}$ and $\sigma_{W}^{\rm EFT2}$ can be factorized as following forms:
\begin{eqnarray}
&&\sigma_i^{\rm int}=\frac{C_i}{\Lambda^4}f_i(M^{\rm min}_{WW},M^{\rm max}_{WW}),\label{eq:fact8int}\\
&&\sigma_{W}^{\rm EFT2}=\frac{C_W^2}{\Lambda^4}f_{\rm EFT2}(M^{\rm min}_{WW},M^{\rm max}_{WW}),
\end{eqnarray}
where $M^{\rm min}_{WW}$ and $M^{\rm max}_{WW}$ are cuts on the diboson invariant mass, 
as a consequence, $|\sigma_i^{\rm int}/\sigma_{W}^{\rm EFT2}|=1$ indicates:
\begin{eqnarray}
\left|\frac{C_i}{C_W^2}\right|=\left|\frac{f_{\rm EFT2}(M^{\rm min}_{WW},M^{\rm max}_{WW})}{f_i(M^{\rm min}_{WW},M^{\rm max}_{WW})}\right|.
\end{eqnarray}
For the fixed $M_{WW}^{\rm min}$, increasing the $M_{WW}^{\rm max}$  saturates the values of both $f_i$ and $f_{\rm EFT2}$ because  the contribution from the high center of mass energy is suppressed by the parton luminosity despite an enhancement in the parton level cross-section due to its scaling with $S$. 
Furthermore, one can also find that the solid lines that cut on a relatively high diboson invariant mass tend to decrease as $\Lambda$ increase, 
this is because the up-type and down-type quark-initiated processes have different coefficients in the $S$ expansion and the scaling behavior for the corresponding parton luminosity functions for $u\bar{u}$ and $d\bar{d}$ are different.
Suppose we parameterize the parton level cross-sections for the dim-8 interference and dim-6 square in the high energy limit in the following forms:
\begin{eqnarray}
&&\hat{\sigma}^{\rm int}_{u} \sim \frac{C_i}{\Lambda^4}k^{\rm  int}_u S,\quad \hat{\sigma}^{\rm int}_{d} \sim \frac{C_i}{\Lambda^4}k^{\rm int}_d S,\\
&&\hat{\sigma}^{\rm EFT2}_{u} \sim \frac{C_W^2}{\Lambda^4}k^{\rm EFT2}_u S,\quad \hat{\sigma}^{\rm EFT2}_{d} \sim \frac{C_W^2}{\Lambda^4}k^{\rm EFT2}_d S,
\end{eqnarray}
where we factor out the Wilson coefficient dependence, and $k$'s are some constants. This time $|\sigma_i^{\rm int}/\sigma_{W}^{\rm EFT2}|=1$ indicates:
\begin{eqnarray}
\left|\frac{C_i}{C_W^2}\right|&&=\left|\frac{\int  dS\ L_{u\bar{u}}(S)k^{\rm EFT2}_uS+L_{d\bar{d}}(S)k^{\rm EFT2}_dS}{\int dS\ L_{u\bar{u}}(S)k^{\rm  int}_uS+L_{d\bar{d}}(S)k^{\rm  int}_dS}\right|,\label{eq:CiCW1}\\
&&=\left|\frac{k^{\rm EFT2}_u}{k^{\rm  int}_u}\left(1+\frac{\int  dS\ L_{d\bar{d}}(S)\left(k^{\rm EFT2}_d/k^{\rm EFT2}_u-k^{\rm  int}_d/k^{\rm  int}_u\right)S}{\int  dS\ L_{u\bar{u}}(S)S+L_{d\bar{d}}(S)k^{\rm  int}_d/k^{\rm  int}_u S}\right)\right|,\label{eq:CiCW}
\end{eqnarray}
where the $L_{u\bar{u}}$ and $L_{d\bar{d}}$ are the parton luminosity functions for $u\bar{u}$ and $d\bar{d}$. The second term in the parenthesis in eq.~\eqref{eq:CiCW} accounts for the dependence of $\Lambda$ for solid lines through the $\Lambda$ dependence in the integral upper limit.

From the plots of the bottom row in figure~\ref{fig:WWEFT2}, one can find that in the high energy regime the dim-8-SM interference cross-section is much larger than dim-6-SM interference cross-section, as a small ratio of $|C_i/C_W|<{\cal O}(1)$ is needed to equate the two cross-sections. This is expected, since at the high energy, the dim-6-SM interference is suppressed by $1/S$, while dim-8-SM interference is enhanced by $S$. 
In contrast, one can find that the dashed lines increase very fast as one increases the NP scale $\Lambda$, this is a manifest result of improving EFT validity by increasing the NP scale. Again one can also understand this feature by factorizing the $\sigma_W^{\rm int}$ into the following form:
\begin{eqnarray}
\sigma_{W}^{\rm int}=\frac{C_W}{\Lambda^2M_W^2}f_{\rm W}(M^{\rm min}_{WW},M^{\rm max}_{WW}),
\end{eqnarray}
where we  extract out an additional factor of $M^2_W$ such that the dimension of $f_W$ matches that of $f_i$ in eq.~\eqref{eq:fact8int}. This time $|\sigma_i^{\rm int}/\sigma_{W}^{\rm int}|=1$ yields:
\begin{eqnarray}
\left|\frac{C_i}{C_W}\right|=\left|\frac{\Lambda^2f_{W}(M^{\rm min}_{WW},M^{\rm max}_{WW})}{M_W^2 f_i(M^{\rm min}_{WW},M^{\rm max}_{WW})}\right|.
\end{eqnarray}
Therefore the  Wilson coefficients ratio increases with $\Lambda^2$, provided that the ratio of $f_{W}$ and $f_i$ saturates for fixed $M_{WW}^{\rm min}=2M_W$ and a large enough $M_{WW}^{\rm max}$.

\subsection{$WZ$ Final State}
In this section, We also examine the operators that contribute to the $WZ$ final state, namely ${\cal O}_{4-7}$, ${\cal O}_{11-13}$, and $O_{18}$. Among these operators, $O_{5,7,11}$ are CP-odd operators, as explained in appendix~\ref{app:cp}, and result in zero interference amplitudes at leading order due to the second non-interference scenario discussed in Section~\ref{sec:nonintclass}. However, unlike in the $WW$ final state, this non-interference effect can be resurrected by including the gauge boson width in propagators, which results in the amplitude after summing over the gauge boson helicities being proportional to $\Gamma_W/M_W$. 
This is because the CP conjugate of the $W^+Z$ final state is $W^-Z$, and the cancellation between the CP conjugate processes by summing over the helicity configurations, as done for the $WW$ final state, does not apply here. On the other hand, $O_{18}$ generates amplitudes only involving the right-handed quark, contrary to the SM, where only the left-handed quark contributes, thus resulting in vanishing interference amplitudes. Table~\ref{tab:scalingWZ} lists the scaling behavior of the interference amplitudes for all the dim-8 operators that contribute to $u\bar{d}\to W^+Z$. The relevant analytical results can be found in~\cite{ampresult}.

\begin{table}[h]
\begin{center}
\renewcommand{\arraystretch}{1.5}
\begin{tabular}{ |c|c|c| } 
 \hline
 Operator & $ 2\operatorname{Re}({\cal A}^{\rm SM}{\cal A}^{\rm NP *})$ & $2\int  d\Omega \operatorname{Re}({\cal A}^{\rm SM}{\cal A}^{\rm NP *})$  \\ 
 \hline
 ${\cal O}_4$ & $u\bar{d}:\  e_4 S^2 +f_4 S+g_4
 $ & $\ov{e}_4 S^2 +\ov{f}_4 S+\ov{g}_4$\\ 
 \hline
 ${\cal O}_5$ & $u\bar{d}:\  \frac{\Gamma_W}{M_W}( f_5 S+g_5)$  & $\frac{\Gamma_W}{M_W}(\ov{f}_5 S+\ov{g}_5)$\\ 
 \hline
 ${\cal O}_6$ & $u\bar{d}:\  e_6 S^2 +f_6 S+g_6$  & $\ov{e}_6 S^2 +\ov{f}_6 S+\ov{g}_6$\\ 
 \hline
 ${\cal O}_7$ & $u\bar{d}:\  \frac{\Gamma_W}{M_W}( f_7 S+g_7)$  & $\frac{\Gamma_W}{M_W}(\ov{f}_7 S+\ov{g}_7)$\\ 
 \hline
 ${\cal O}_{11}$ & $u\bar{d}:\  g_{11}\frac{\Gamma_W}{M_W}$  & 0\\ 
 \hline
 ${\cal O}_{12}$ & $u\bar{d}:\   e_{12} S^2 +f_{12} S+g_{12}$  & $\ov{e}_{12} S^2 +\ov{f}_{12} S+\ov{g}_{12}$\\ 
 \hline
 ${\cal O}_{13}$ & $u\bar{d}:\   e_{13} S^2 +f_{13} S+g_{13}$  & $\ov{e}_{13} S^2 +\ov{f}_{13} S+\ov{g}_{13}$\\ 
 \hline
 ${\cal O}_{18}$ & $u\bar{d}:\  0$  & 0\\ 
 \hline
\end{tabular}\caption{Scaling of $q\bar{q}\to WZ$ interference amplitude.}\label{tab:scalingWZ}
\end{center}
\end{table}

\begin{figure}[htb]
\centering
\includegraphics[width=0.49\textwidth]{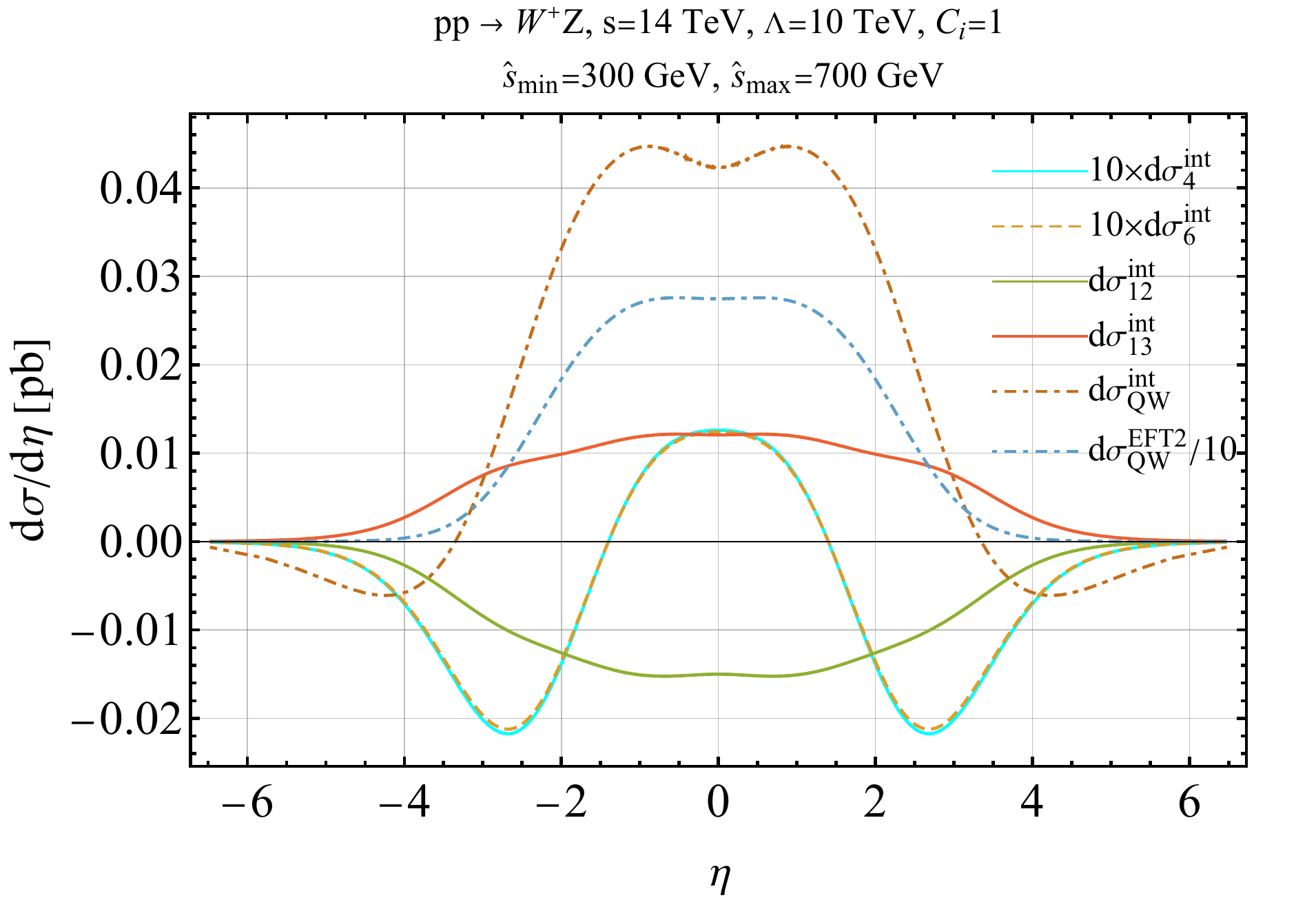}
\caption{Differential cross-section for dimension-8 and dimension-6 operators for $pp\to W^+Z$. The dotted cyan line and dashed yellow line for ${\cal O}_4$ and ${\cal O}_6$ respectively are almost overlapped with each other, they are also scaled by a factor of 10. The brown and blue dash-dotted lines are  the ${\rm BSM}\times {\rm SM}$ interference and the ${\rm BSM}^2$ contributions for the dimension-6 ${\cal O}_{W}$ respectively.}\label{fig:dsigWZ}
\end{figure}

In figure~\ref{fig:dsigWZ}, we present the hadronic angular distribution for the $W^+$ boson in the production process $pp\to W^+ Z$, where $\theta$ is defined as the angle between the $W^+$ and the beam axis in the diboson frame. It can be observed that the differential interference cross-sections of ${\cal O}{4}$ and ${\cal O}{6}$ are smaller compared to those of ${\cal O}{12}$ and ${\cal O}{13}$. Furthermore, the angular distributions for ${\cal O}{4}$ and ${\cal O}{6}$ are highly similar, as can be seen from the plot where they overlap with each other to a significant extent, making it difficult to differentiate between the effects of these two operators in the $WZ$ production channel based on this distribution alone.

\begin{figure}
\centering
         \includegraphics[width=0.24\textwidth]{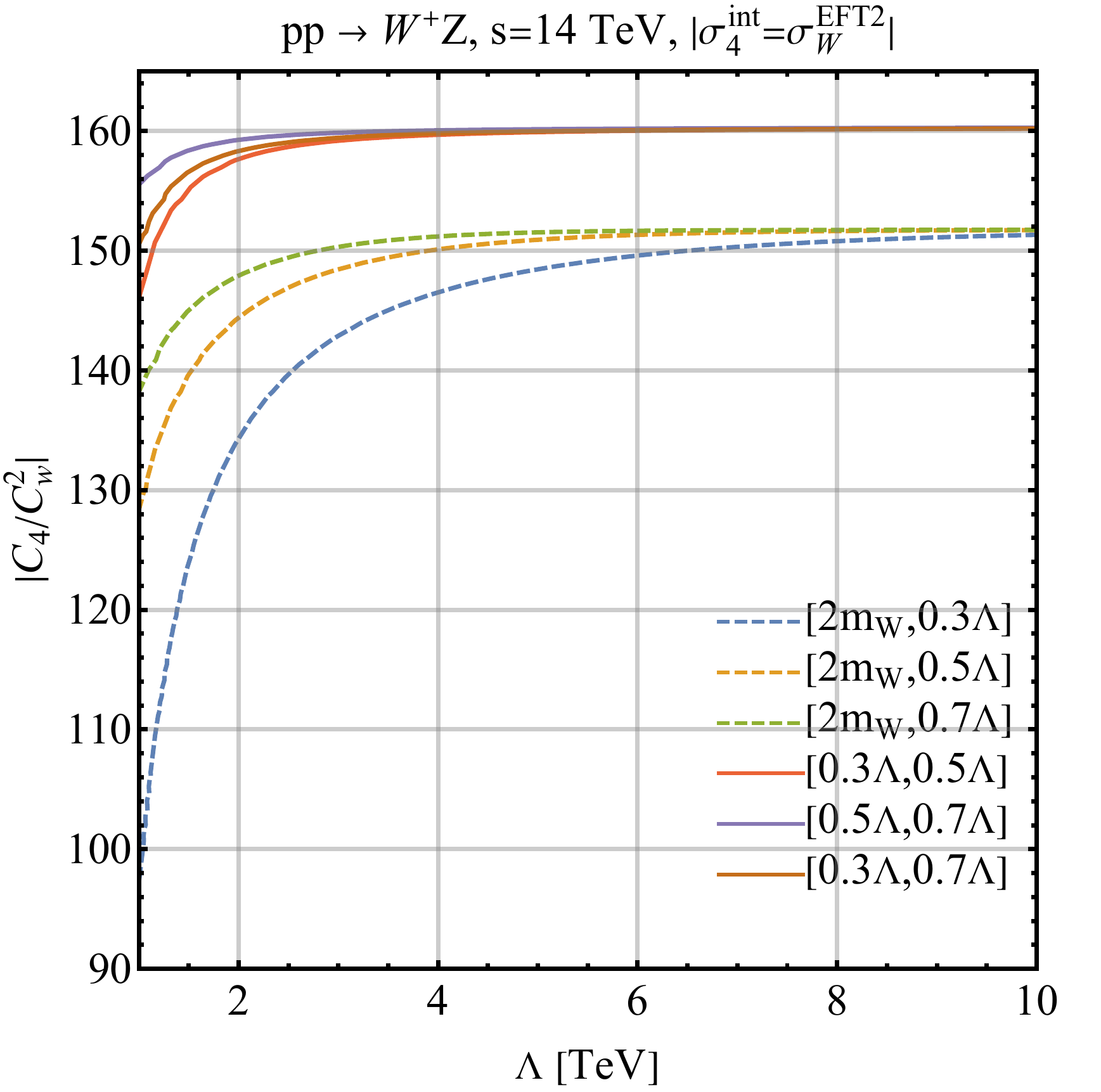}
         \includegraphics[width=0.24\textwidth]{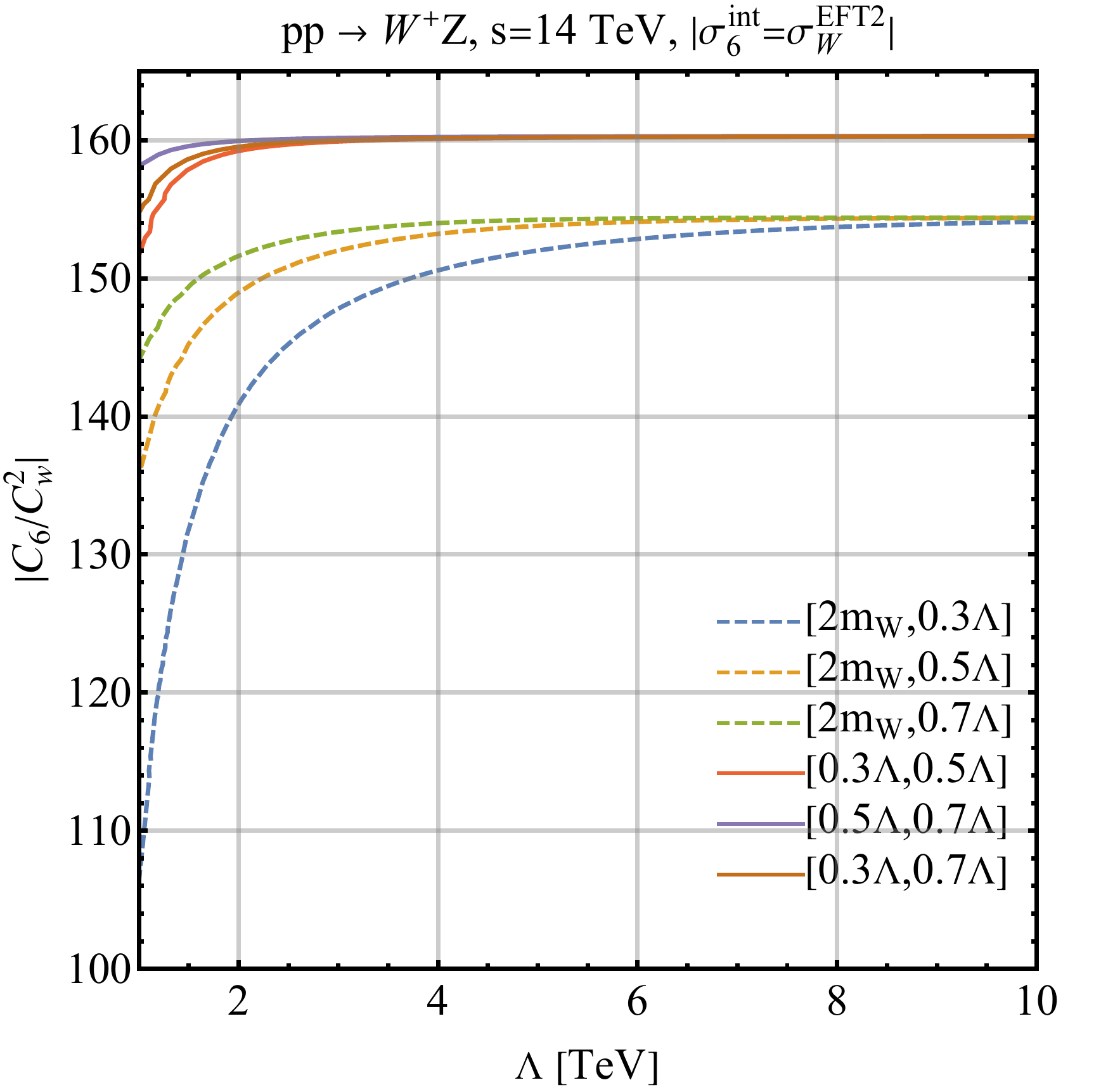}
         \includegraphics[width=0.24\textwidth]{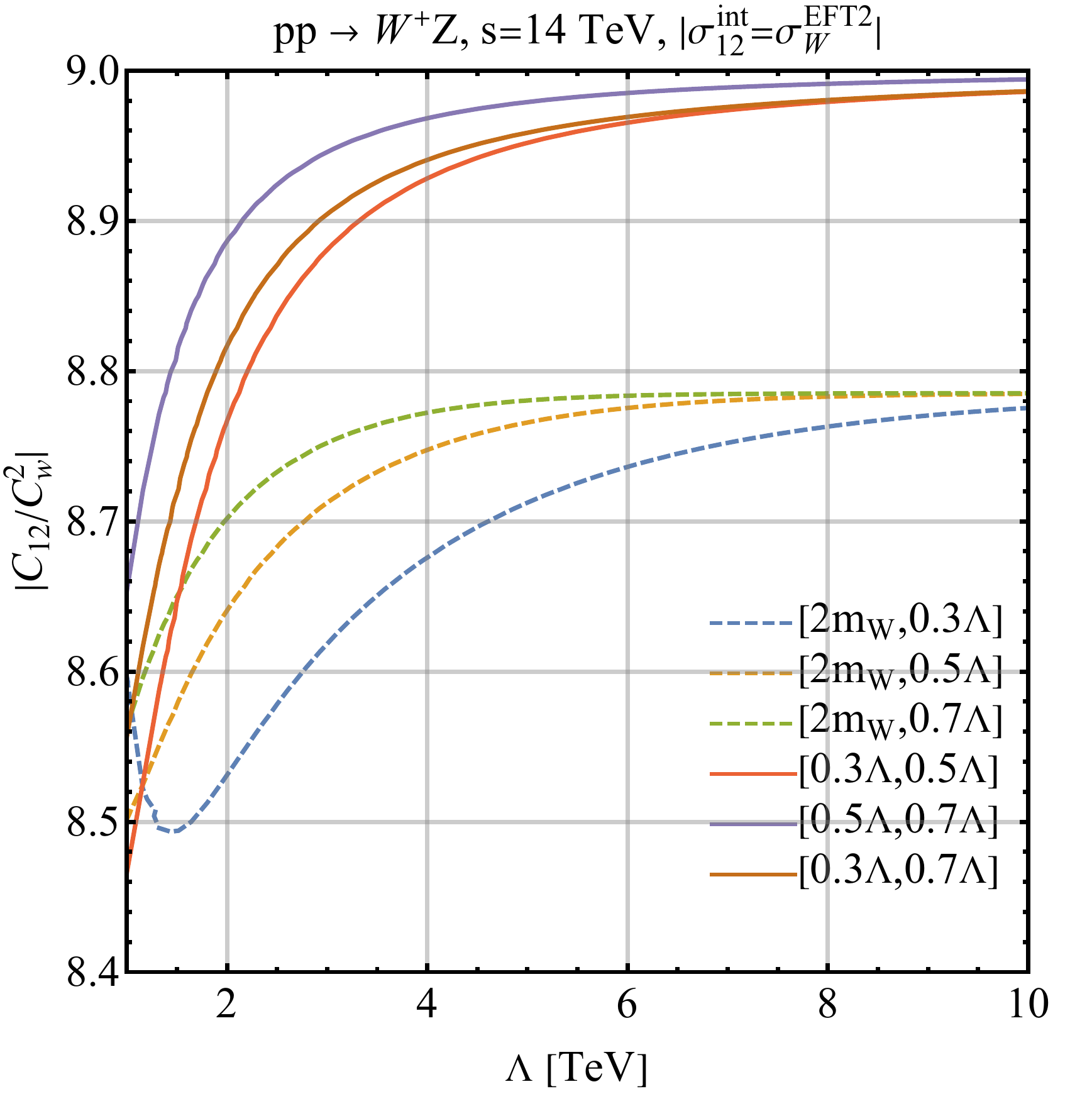}
         \includegraphics[width=0.24\textwidth]{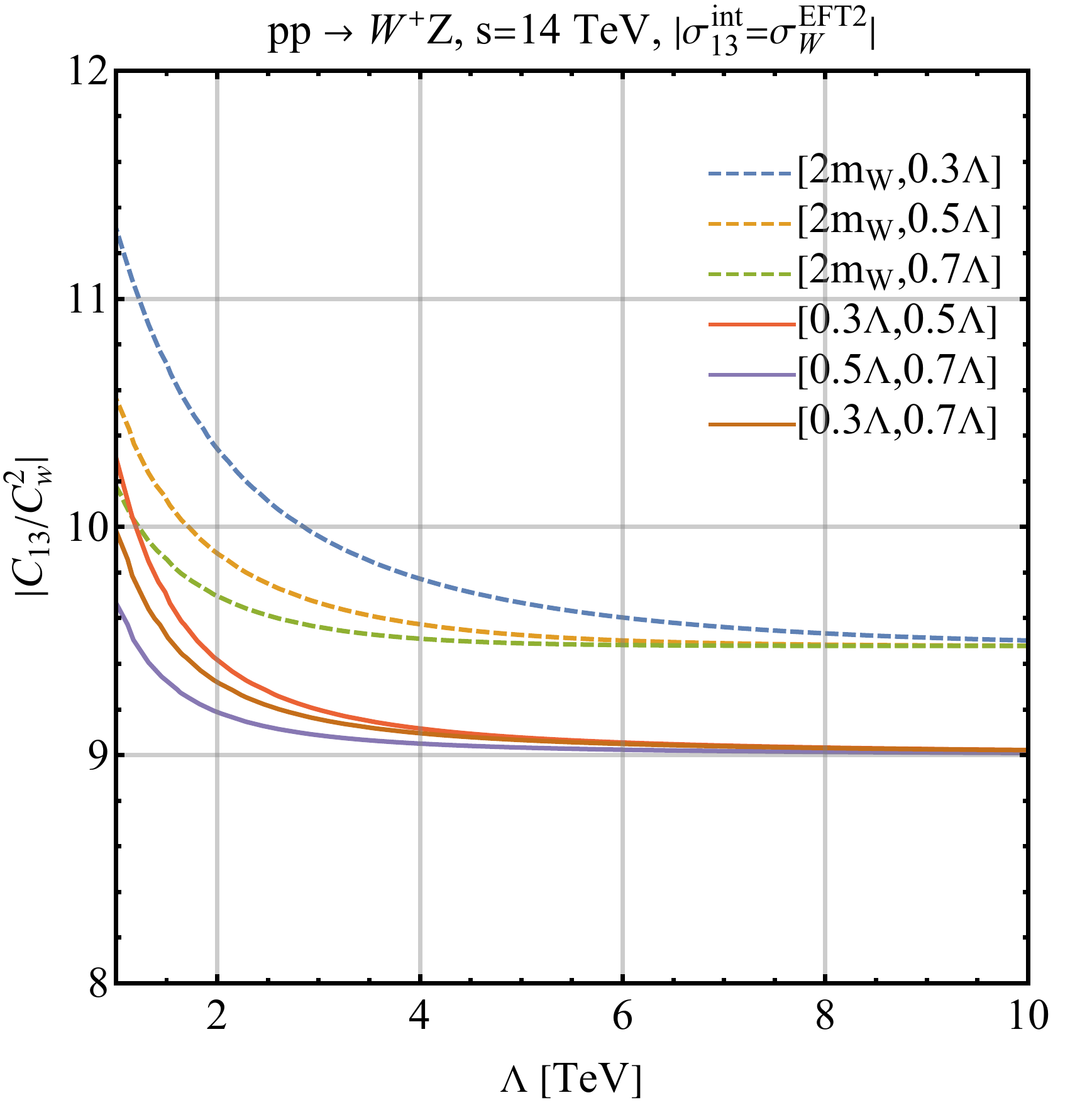}
         \includegraphics[width=0.24\textwidth]{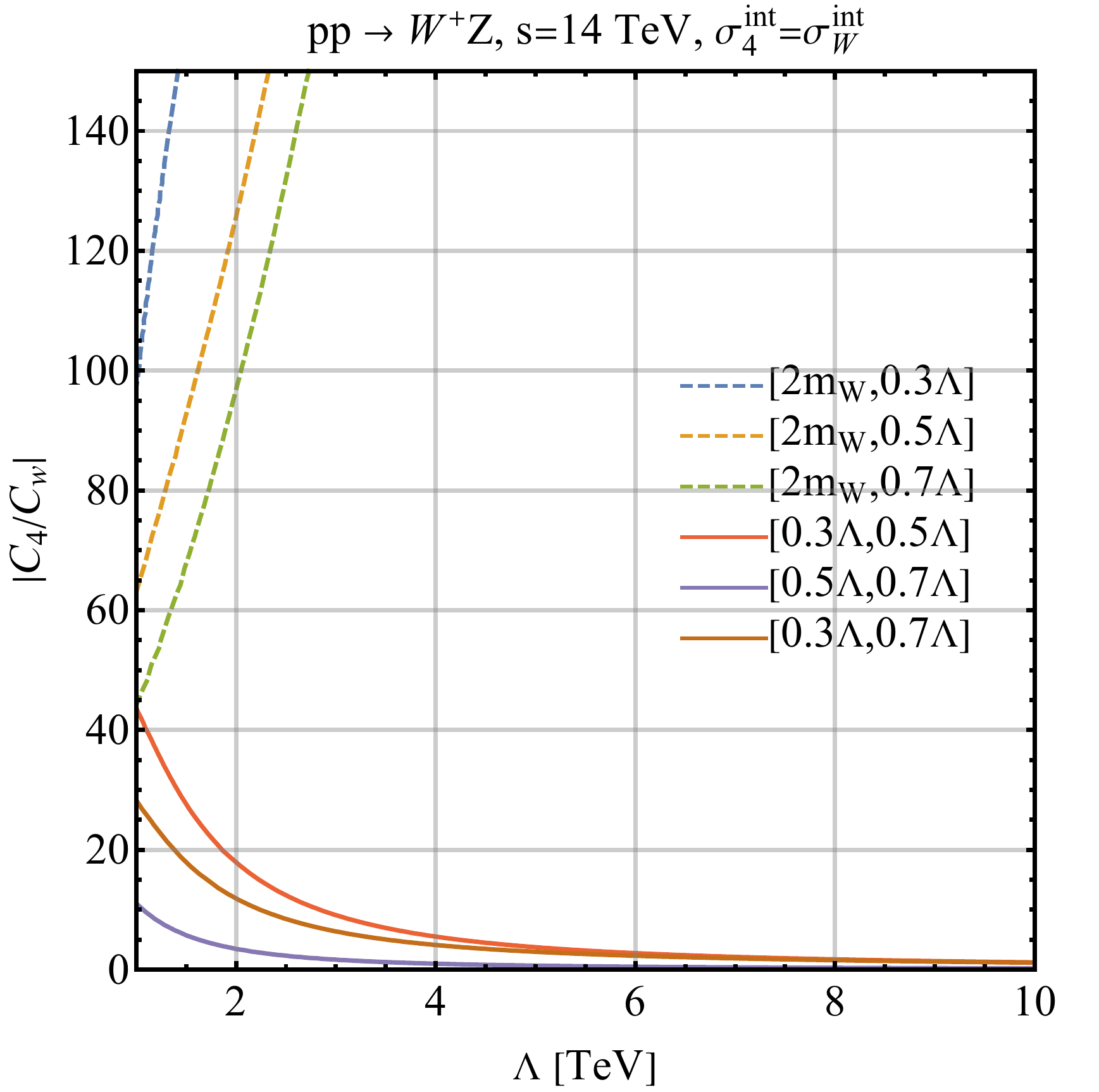}
         \includegraphics[width=0.24\textwidth]{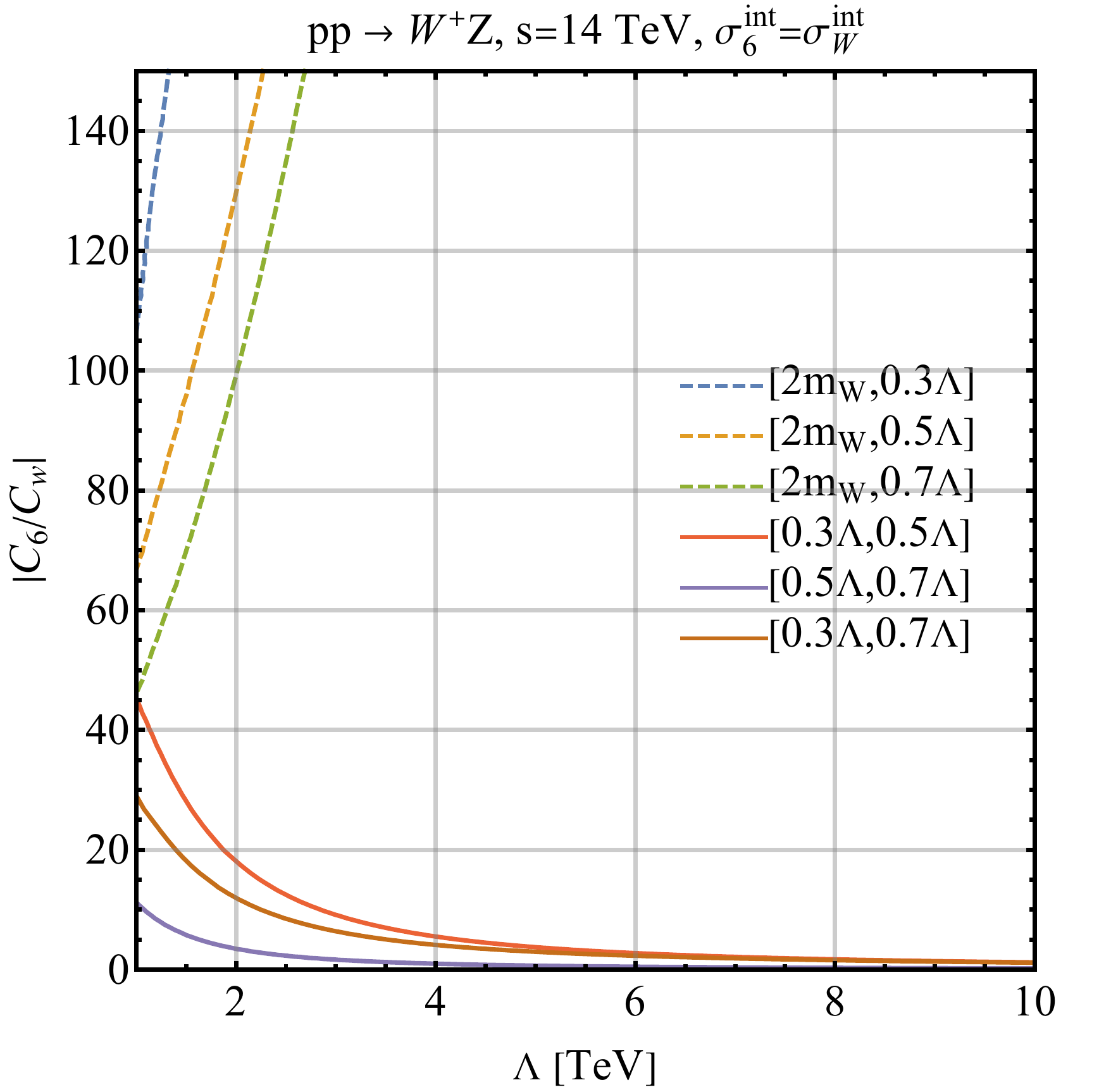}
         \includegraphics[width=0.24\textwidth]{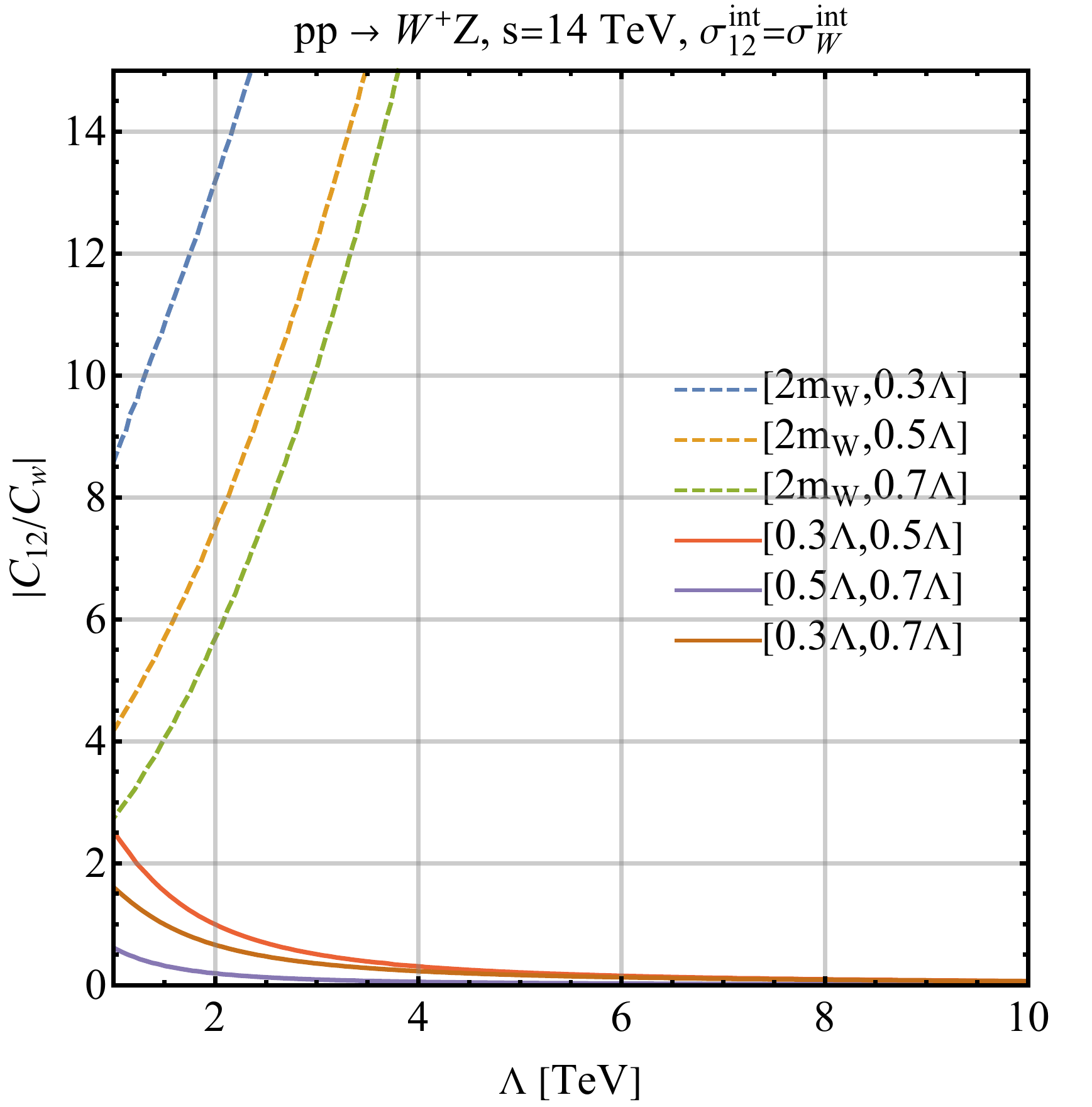}
         \includegraphics[width=0.24\textwidth]{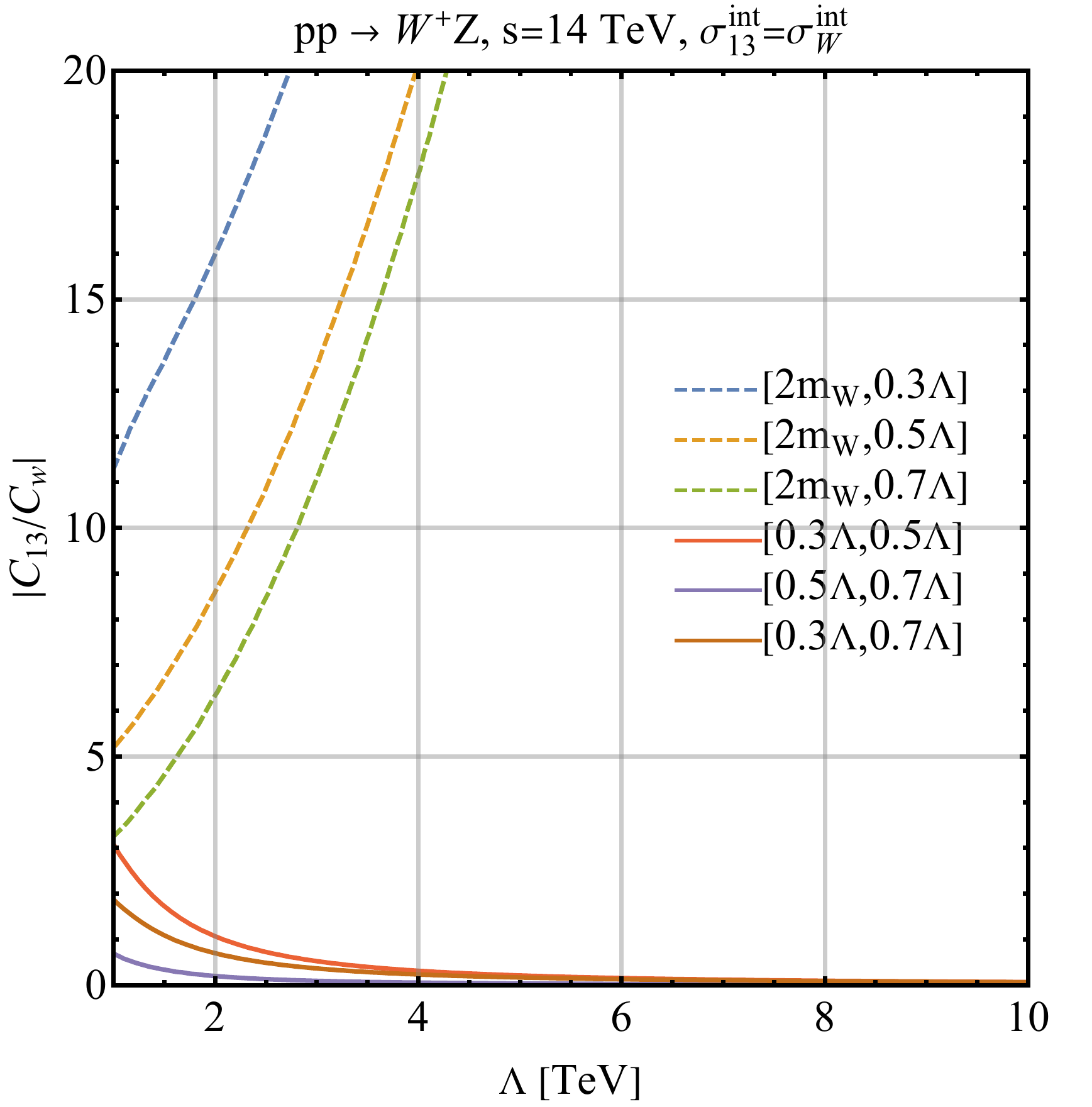}
\caption{Comparisons of interference cross-sections of the dim-8 operator ${\cal O}_{4}$, ${\cal O}_{6}$, ${\cal O}_{12}$ and ${\cal O}_{13}$ respectively from left to right with respect to the cross-section from the dim-6 amplitude square (top) and dim-6 interference (bottom) for $pp\to W^+Z$. Different colors represent different cuts on the diboson invariant mass. Each line represents the ratio of the dim-8 and dim-6 Wilson coefficients such that the dim-8 interference cross-section and the dim-6 square cross-section (top) or dim-6 interference cross-section (bottom) have the same magnitude.  }\label{fig:WZEFT2}
\end{figure}

In the first row of figure~\ref{fig:WZEFT2}, we plot the ratio of the dim-8 dimensionless Wilson coefficient $C_i$ to $C_W$ against the NP scale $\Lambda$, ensuring that the dim-8 interference cross-sections are equal to the dim-6 square (dim-6 interference) cross-sections. The colors in the plot represent various cuts on the invariant mass of the diboson system. In contrast to the $WW$ final state, we observe that the solid lines, which pertain to the high invariant mass region, approach a constant. This behavior arises because the $WZ$ final state only involves the $u\bar{d}$ initiated process (ignoring the $c\bar{s}$ initiated process). Therefore, only terms proportional to $L_{u\bar{d}}\cdot S$ exist within the integral in both numerators and the denominator of equation~\eqref{eq:CiCW1}, canceling out with each other and leaving a constant ratio.

In the second row of figure~\ref{fig:WZEFT2}, we compare the dim-8 and dim-6 interference cross-sections. Similar to the $WW$ final state, the increasing dashed line indicates an improvement in the convergence of the EFT as the NP scale increases. The decreasing solid lines reflect that at high energy, the interference effect of dim-8 dominates over that of dim-6 due to the faster growth of the dim-8 interference amplitude with increasing energy.

\section{Conclusion and Outlook}\label{sec:conclude}
We investigate the dim-8 SMEFT contributions to diboson production with $WW/WZ$ final states. We classify all the operators that are capable of generating $E^4/\Lambda^4$ enhancement in the interference amplitude, and quantitatively compare their values with the dim-6 interference and dim-6 square amplitudes. 
We find that the non-interference found for dimension-six operators is not observed for dimension-eight and therefore further limit the legitimacy of constraining operators mainly from the square of the dimension-six amplitudes. Preliminary result seems to indicate that this is also true for the other electroweak  diboson processes. 
We find that the dim-8 operator ${\cal O}_{10}$ can generate the interference cross-section for the process $pp\to W^+W^-$ that is comparable with the dim-6 interference cross-section from the operator ${\cal O}_W$ assuming equal dimensionless Wilson coefficients. For the $WZ$ final state, the interference cross-section generated by ${\cal O}_{12}$ or ${\cal O}_{13}$ is also comparable with the dim-6 interference cross-section from ${\cal O}_W$. 
We also point out that at the total cross-section level, the non-interference can occur due to the selection rule of the angular momentum of the scattering process, while suppression of the interference cross-section can also result from the cancellation between the $q\bar{q}$ and $\bar{q}q$ initiated channels in the pp collider.
Both of these effects can be analyzed with the J-basis operators prior to the derivation of Feynman rules and the calculation of amplitudes. 

Finally, we note that precise bounds on the Wilson coefficients of various operators require dedicated collider simulations. We anticipate that the semileptonic decay channel, where one of the vector bosons decays leptonicly, will have better sensitivity in the moderate invariant mass regime. This is due to the lepton signature which can effectively reduce the large QCD background. For the higher invariant mass regime, we expect the sensitivity for fully hadronic decay to be better by reconstructing two opposite fat jets and a larger hadronic decay branching ratio. It is also important to consider detector effects in determining the extent to which we can exploit the information in the angular distribution shown in figure~\ref{fig:dsigWW} and \ref{fig:dsigWZ} to enhance the signal sensitivity and discerning effects from different operators. Furthermore, when considering the decay, additional dim-8 operators such as $\bar{\psi}\psi\bar{\psi}\psi F^{\mu\nu}$ can interfere with the SM full amplitude. We acknowledge that further studies are needed to address these issues.

\acknowledgments

HL thanks Rafael Aoude, Olivier Mattelaer, Zhe Ren, Ming-Lei Xiao and Yu-Hui Zheng  for the helpful discussion. HL is supported by the 4.4517.08 IISN-F.N.R.S convention.

\appendix

\section{Spinor Conventions}
\label{app:spinor}
4-component Dirac field $\Psi$ can be expressed in terms of two 2-component left-handed Weyl fermions $\xi$ and $\chi$ by:
\begin{align}
	\Psi=\left(\begin{array}{c} \xi_{\alpha}\\\chi^{\dagger\dot{\alpha}} \end{array} \right),\quad \bar{\Psi}=\Psi^{\dagger}\gamma^0=\left(\chi^{\alpha},\;\xi^{\dagger}_{\dot{\alpha}} \right)\;.
\end{align}
Conventions between the 2-component fields and 4-component field in the SM are summarized in the following formulae:
\begin{align}
	q_{\rm{L}}=\begin{pmatrix}Q\\0\end{pmatrix},\quad u_{\rm{R}}=\left(\begin{array}{c}0\\u_{_\mathbb{C}}^{\dagger}\end{array}\right),\quad d_{\rm R}=\left(\begin{array}{c}0\\d_{_\mathbb{C}}^{\dagger}\end{array}\right),\quad l_{\rm L}=\left(\begin{array}{c}L\\0\end{array}\right),\quad e_{\rm R}=\left(\begin{array}{c}0\\e_{_\mathbb{C}}^{\dagger}\end{array}\right).\\
	\bar{q}_{\rm{L}}=\left(0\,,\,Q^{\dagger} \right),\quad \bar{u}_{\rm{R}}=\left(u_{_\mathbb{C}}\,,\,0 \right),\quad \bar{d}_{\rm R}=\left(d_{_\mathbb{C}}\,,\,0\right),\quad \bar{l}_{\rm L}=\left(0\,,\,L^{\dagger}\right),\quad \bar{e}_{\rm R}=\left(e_{_\mathbb{C}}\,,\,0\right).
\end{align}
$\sigma^\mu$ and $\bar{\sigma}^\mu$ is defined by
$\sigma^{\mu}_{\alpha\dot{\alpha}}=(\mathbbm{1}_{\alpha\dot{\alpha}},{\boldsymbol\sigma}^i_{\alpha\dot{\alpha}})^\mu$, $\bar{\sigma}^{\mu\dot{\alpha}\alpha}=(\mathbbm{1} ^{\dot\alpha\alpha},-{\boldsymbol\sigma}^{i\dot{\alpha}\alpha})^\mu$, with $\boldsymbol\sigma^i$ ordinary Pauli matrices. We have the relation between $\sigma^\mu$ and $\bar{\sigma}^\mu$:
\begin{equation}
    \bar\sigma^{\mu\dot\alpha\alpha}=\epsilon^{\alpha\beta}\epsilon^{\dot{\alpha}\dot{\beta}}\sigma^{\mu}_{\beta\dot{\beta}}.
\end{equation}
where the Levi-Civita tensors have conventions $\epsilon^{12}=\epsilon_{21}=+1$.  Charge conjugate matrix $C=i\gamma^0\gamma^2$, Dirac matrices $\gamma^\mu$ and the Lorentz generator $\sigma^{\mu\nu}$ have following forms in the chiral representation:
\begin{eqnarray}
&&C=\begin{pmatrix} \epsilon_{\alpha\beta}&0\\0&\epsilon^{\dot{\alpha}\dot{\beta}}\end{pmatrix}=\begin{pmatrix} -\epsilon^{\alpha\beta}&0\\0&-\epsilon_{\dot{\alpha}\dot{\beta}}\end{pmatrix},\quad
\gamma^{\mu}=\begin{pmatrix}
0&\sigma^{\mu}_{\alpha\dot{\beta}}\\\bar{\sigma}^{\mu\dot{\alpha}\beta}&0
\end{pmatrix},\nonumber \\
&&\sigma^{\mu\nu}=\dfrac{i}{2}[\gamma^\mu,\gamma^\nu]=\begin{pmatrix}
\left(\sigma^{\mu\nu}\right)_\alpha{}^\beta&0\\0&\left(\bar{\sigma}^{\mu\nu}\right)^{\dot{\alpha}}{}_{\dot{\beta}}
\end{pmatrix}
\end{eqnarray}
We have the following useful identities changing the $\bar{\sigma}$ to $\sigma$ in the fermion current:
\begin{eqnarray}
\xi^\dagger_{1\dot\alpha}\bar{\sigma}^{\mu\dot{\alpha}\alpha}\xi_{2\alpha}=-\xi_{2}^{\alpha}\sigma^\mu_{\alpha\dot\alpha}\xi^{\dagger\dot\alpha}.
\end{eqnarray}
We also list the conversion between two-component and four-component spinors present in the Feynman rule or scattering amplitude in table~\ref{tab:multicol}. Various definitions related to spinors are defined as follows.
We first introduce two eigenvectors of the helicity operators $\hat{h}=\mathbf{p}\cdot \boldsymbol{\sigma}/|\mathbf{p}|$ for momentum $\mathbf{p}$ in the direction $\hat{\mathbf{p}} = (\sin\theta\cos\phi, \sin\theta\sin\phi, \cos\theta)$:
\begin{eqnarray}
 &&  \chi^{-}_\alpha=\begin{pmatrix}
-e^{-i\phi}\sin(\theta/2)\\
\cos(\theta/2)
\end{pmatrix},\quad
\chi^{+\dot\alpha}=\begin{pmatrix}
\cos(\theta/2)\\
e^{i\phi}\sin(\theta/2)
\end{pmatrix}, 
\end{eqnarray} 
with $\pm$ indicating their eigenvalue of the helicity operator.
then we fix the momentum dependence of the massless spinor variables to the following forms:
\begin{eqnarray}
\lambda_\alpha=\sqrt{2E}\chi^{-}_\alpha,\quad
\tilde{\lambda}^{\dot\alpha}=\sqrt{2E}\chi^{+\dot\alpha}.
\end{eqnarray}
One can easily verify that:
\begin{eqnarray}
    (\lambda^\alpha)^* = \tilde{\lambda}^{\dot\alpha} = \epsilon^{\dot{\alpha}\dot{\beta}}\tilde{\lambda}_{\dot\beta},\quad  (\tilde\lambda_{\dot\alpha})^* = {\lambda}_{\alpha} = \epsilon_{\dot{\alpha}\dot{\beta}}{\lambda}^{\beta}. 
\end{eqnarray}
We also define the helicity eigenstates of the massive Dirac 4-component spinors in the following forms:
\begin{eqnarray}
    u^{h}(p) =\begin{pmatrix}
        \sqrt{E-h|\mathbf{p}|}\chi^{(h)}\\
        \sqrt{E+h|\mathbf{p}|}\chi^{(h)}
    \end{pmatrix} ,\quad
        v^{h}(p) =\begin{pmatrix}
        -\sqrt{E+h|\mathbf{p}|}\chi^{(-h)}\\
        \sqrt{E-h|\mathbf{p}|}\chi^{(-h)}
    \end{pmatrix} 
\end{eqnarray}
\begin{table}[ht]
\caption{Relations between 2- and 4-component notation for massless fermion}
\begin{center}
\begin{tabular}{c|c|c|c}
    \hline
    fermion type & state type & helicity & notaion\\
    \hline
    \multirow{4}{*}{Particle}&\multirow{2}{*}{In-coming}&$+$&${\rm P}_{\rm R}u=u^+=\tilde{\lambda}^{\dot\alpha}$, $ \overline{v}{\rm P}_{\rm R}=\overline{v^+}=\tilde{\lambda}_{\dot\alpha}$\\
    \cline{3-4}
    &&$-$&${\rm P}_{\rm L}u=u^-={\lambda}_{\alpha}$, $ \overline{v}{\rm P}_{\rm L}=\overline{v^-}={\lambda}^{\alpha}$\\
    \cline{2-4}
    &\multirow{2}{*}{Out-going}&$+$&$\overline{u}{\rm P}_{\rm L}=\overline{u^+}={\lambda}^{\alpha}$, ${\rm P}_{\rm L}v={v^+}=-{\lambda}_{\alpha}$\\
    \cline{3-4}
    &&$-$&$\overline{u}{\rm P}_{\rm R}=\overline{u^-}=\tilde{\lambda}_{\dot\alpha}$, ${\rm P}_{\rm R}v= {v^-}=\tilde{\lambda}^{\dot\alpha}$\\
    \hline
    \multirow{4}{*}{Anti-Particle}&\multirow{2}{*}{In-coming}&$+$&$\overline{v}{\rm P}_{\rm R}=\overline{v^+}=\tilde{\lambda}_{\dot\alpha}$, ${\rm P}_{\rm R}u= {u^+}=\tilde{\lambda}^{\dot\alpha}$\\
    \cline{3-4}
    &&$-$&$\overline{v}{\rm P}_{\rm L}=\overline{v^-}= {\lambda}^{ \alpha}$, ${\rm P}_{\rm L}u= {u^-}={\lambda}_{\alpha}$\\
    \cline{2-4}
    &\multirow{2}{*}{Out-going}&$+$&${\rm P}_{\rm L}v={v^+}={-\lambda}_{\alpha}$, $\overline{u}{\rm P}_{\rm L}=\overline{u^+}={\lambda}^{\alpha}$\\
    \cline{3-4}
    &&$-$&${\rm P}_{\rm R}v={v^-}=\tilde{\lambda}^{\dot\alpha}$, $\overline{u}{\rm P}_{\rm R}=\overline{u^-}=\tilde{\lambda}_{\dot\alpha}$\\
    \hline
\end{tabular}
\end{center}
\label{tab:multicol}
\end{table}

\section{CP Properties of operators}\label{app:cp}
Following the convention in  eq.(13.9) and eq.(13.10) in Ref.~\cite{Branco:1999fs}, the CP transformation of the $W$ and $Z$  bosons can be expressed as:
\begin{eqnarray}
&&({\cal CP})W^{+\mu}(t,\mathbf{r})({\cal CP})^\dagger = -e^{i\xi_W}W^-_\mu(t,-\mathbf{r})\\
&&({\cal CP})W^{-\mu}(t,\mathbf{r})({\cal CP})^\dagger = -e^{-i\xi_W}W^+_\mu(t,-\mathbf{r})\\
&&({\cal CP})Z^{\mu}(t,\mathbf{r})({\cal CP})^\dagger = -Z_\mu(t,-\mathbf{r}),
\end{eqnarray}
from which by setting the artibrary phase $\xi_W=0$ one can deduce that:
\begin{eqnarray}
&&({\cal CP})W^{1\mu }(t,\mathbf{r})({\cal CP})^\dagger=- W_\mu^1(t,-\mathbf{r})\\
&&({\cal CP})W^{2\mu }(t,\mathbf{r})({\cal CP})^\dagger= W_\mu^2(t,-\mathbf{r})\\
&&({\cal CP})W^{3\mu }(t,\mathbf{r})({\cal CP})^\dagger=- W_\mu^3(t,-\mathbf{r}), 
\end{eqnarray}
which can be collectively expressed as $({\cal CP})W^{I\mu }(t,\mathbf{r})({\cal CP})^\dagger=- s^I W_\mu^I(t,-\mathbf{r})$ with $s^{1,3}=1,\ s^{2}=-1$.
Under such a convention one can derive the CP transformation properties of the field strength tensor of the $SU(2)$ gauge boson $W_{\mu\nu}^I$:
\begin{eqnarray}
W_{\mu\nu}^I = \partial_\mu W_{\nu}^I-\partial_\nu W_{\mu}^I+g\epsilon^{IJK}W_{\mu}^JW_{\nu}^K&&,\quad \tilde{W}_{\mu\nu}^I =\frac{1}{2}\epsilon_{\mu\nu\rho\sigma}W^{I\rho\sigma}\\
({\cal CP})\partial_\mu W_{\nu}^I(t,\mathbf{r})({\cal CP})^\dagger &&= -s^{I}[\partial^\mu W^{I\nu}](t,-\mathbf{r})\\
({\cal CP})\epsilon^{IJK}W_{\mu}^JW_{\nu}^K(t,\mathbf{r})({\cal CP})^\dagger &&= \epsilon^{IJK}s^{J}s^{K}[W^{J\mu }W^{K\nu }](t,-\mathbf{r})\nonumber \\
&&=-s^I\epsilon^{IJK}[W^{J\mu }W^{K \nu }](t,-\mathbf{r})\label{eq:WCP2}\\
({\cal CP})  W_{\mu\nu}^I(t,\mathbf{r})({\cal CP})^\dagger &&=-s^IW^{I\mu\nu} (t,-\mathbf{r}).\label{eq:WCP3}
\end{eqnarray}
In the above equations, the square bracket on the right-hand side reminds that the derivative operator with upper indices on the right should reads $\partial^\mu[t,-\mathbf{r}]= (\partial^t, \partial^{-\mathbf{r}})$ which is equivalent to the derivative $ \partial_\mu[t,\mathbf{r}]$ on the left with lower indices, that is the operator $({\cal CP})$ only acts on the fields. In eq.~\eqref{eq:WCP2}, one can easily verify that $\epsilon^{IJK}s^{J}s^{K}= -s^I\epsilon^{IJK}$, without summation of the repeated indices. Similarly, the CP transformation property of the dual field strength tensor $\tilde{W}_{\mu\nu}^I$ can also be easily derived with eq.~\eqref{eq:WCP3}, we work on the situations $(\mu=0,\nu=i)$ and $(\mu=i,\nu=j )$ with $i,j=(1,2,3)$ separately and verify that they lead to a coherent expression:
\begin{eqnarray}
({\cal CP})  \tilde{W}_{0i}^I(t,\mathbf{r})({\cal CP})^\dagger &&= \frac{1}{2}\epsilon_{0ijk}({\cal CP})  W_{ji}^I(t,\mathbf{r})({\cal CP})^\dagger\nonumber \\
&&=\frac{(-s^I)}{2}\epsilon_{0ijk}\tilde{W}^{Ijk}(t,-\mathbf{r})\nonumber\\
&&=\frac{(s^I)}{2}\epsilon^{0ijk}W^{Ijk}(t,-\mathbf{r})=s^I\tilde{W}^{I0i}(t,-\mathbf{r})\\
({\cal CP})  \tilde{W}_{ij}^I(t,\mathbf{r})({\cal CP})^\dagger &&= \epsilon_{ij0k}({\cal CP})  W_{0k}^I(t,\mathbf{r})({\cal CP})^\dagger\nonumber \\
&&=(-s^I)\epsilon_{ij0k}W^{I0k}(t,-\mathbf{r})\nonumber\\
&&=(-s^I)\epsilon^{ij0k}W^{I0k}(t,-\mathbf{r})=s^I\tilde{W}^{Iij}(t,-\mathbf{r})\\
({\cal CP})  \tilde{W}_{\mu\nu}^I(t,\mathbf{r})({\cal CP})^\dagger &&= s^I\tilde{W}_{\mu\nu}^I(t,-\mathbf{r}).\label{eq:WtCP}
\end{eqnarray}
We can check that the dimension 6 SMEFT operator ${\cal O}_{\tilde{W}}=\epsilon^{IJK}W^IW^J\tilde{W}^K$ is indeed a CP violating operator:
\begin{eqnarray}
({\cal CP}) \epsilon^{IJK}W^IW^J\tilde{W}^K({\cal CP})^\dagger&&=\epsilon^{IJK}(-s^IW^I)(-s^JW^J)(s^K\tilde{W}^K)\nn \\
&&=-s^I\epsilon^{IJK}s^I(W^I)(W^J)(\tilde{W}^K)\nn \\
&&=-\epsilon^{IJK}W^IW^J\tilde{W}^K.\label{eq:CPWWWt}
\end{eqnarray}
For the fermion part, we also adopt the convention in Ref.~\cite{Branco:1999fs}, where under the CP transformation, the fermion bi-linears transform as follows:
\begin{eqnarray}
&&({\cal CP})  \overline{\psi_L}\gamma^\mu\chi_{L}({\cal CP})^\dagger = -e^{i(\xi_\psi-\xi_\chi)}\overline{\psi_L}\gamma_\mu\chi_{L}\label{eq:CPfL}\\
&&({\cal CP})  \overline{\psi_R}\gamma^\mu\chi_{R}({\cal CP})^\dagger = -e^{i(\xi_\psi-\xi_\chi)}\overline{\psi_R}\gamma_\mu\chi_{R}.\label{eq:CPfR}
\end{eqnarray}
If we consider the flavor diagonal effective operators, then  $\chi=\psi$, and the phase factor on the right-hand side vanish. With these conventions, we can analyse the CP transformation properties of the dimesnion 8 operators ${\cal O}_{1\sim 18}$. As we have already obtained the CP transformation properties of the field strength tensors of the gauge fields in eq.~\eqref{eq:WCP3} and \eqref{eq:WtCP}, we only left with the same analysis for the fermion tensors. 
To proceed, we need to first specify the meaning of the covariant with double arrow present in the fermion tensors as follows:
\begin{eqnarray}
\overline{\psi}_i\gamma^\mu \overleftrightarrow{D}^\nu \chi_j&&\equiv \overline{\psi}_i\gamma^\mu (D^\nu \chi_j)-\overline{D^\nu\chi}_j\gamma^\mu \psi_i,\label{eq:doubleD}\\
\text{with } &&\overline{D^\nu\chi}_j\equiv (D^\nu\chi_j)^\dagger \gamma^0\nonumber\\
&&=(\partial^\nu\chi_j+igT^I_{jk}A^{I\nu}\chi_k)^\dagger\gamma^0\nonumber\\
&&=(\partial^\nu\overline{\chi}_j-ig\overline{\chi}_kT^I_{kj}A^{I\nu}),
\end{eqnarray}
where $i,j,k$ are the gauge indices of certain gauge group, and $T^I$ and $A^I$ are the generators and the gauge field of the corresponding gauge group respectively. Therefore the CP conjugate of the first term on the right-hand side of eq.~\eqref{eq:doubleD} gives:
\begin{eqnarray}
({\cal CP})  \overline{\psi}_i\gamma^\mu (D^\nu \chi_j)({\cal CP})^\dagger&&=({\cal CP})\overline{\psi}_i\gamma^\mu (\partial^\nu \chi_j)({\cal CP})^\dagger+igT^I_{jk}({\cal CP}) \overline{\psi}_i\gamma^\mu A^{I\nu} \chi_k({\cal CP})^\dagger\nonumber \\
&&=-\partial_\nu\overline{\chi}_j\gamma_\mu\psi_i-igT^I_{jk}\overline{\psi}_j\gamma_\mu (-s^I A^{I}_{\nu}) \psi_i\nonumber \\
&&=-\partial_\nu\overline{\chi}_k\gamma_\mu\psi_i-igs^IT^I_{kj}\overline{\chi}_k\gamma_\mu (-s^I A^{I}_{\nu}) \psi_i\nonumber \\
&&=-\partial_\nu\overline{\chi}_k\gamma_\mu\psi_i+igT^I_{kj}\overline{\chi}_k\gamma_\mu  A^{I}_{\nu} \psi_i=-\overline{D_\nu\chi}_j\gamma_\mu \psi_i,\label{eq:CPXDY}
\end{eqnarray}
where in the second line we have used eq.~\eqref{eq:CPfL} and  \eqref{eq:CPfR} ignoring  additional phases, 
while for the gauge field $A^I$ we have $({\cal CP})A^{I\mu }(t,\mathbf{r})({\cal CP})^\dagger=- s^I A_\mu^I(t,-\mathbf{r})$ with $s^I=1$ for $T^{I*}=(T^{I})^{\rm T}=T^I$, and $s^I=-1$ for $T^{I*}=(T^{I})^{\rm T}=-T^I$, which is valid for the $SU(2)$ and $SU(3)$ gauge group we considered in the SMEFT. From the second line to the third line, we have used $T_{jk}^I=s^I T_{jk}^I$. Similar to eq.~\eqref{eq:CPXDY}, one can derive the CP conjugation of $\overline{D^\nu\chi}_j\gamma^\mu \psi_i$, and they together gives:
\begin{eqnarray}
({\cal CP})  \overline{\psi}_i\gamma^\mu \overleftrightarrow{D}^\nu \chi_j({\cal CP})^\dagger = \overline{\chi}_j\gamma^\mu \overleftrightarrow{D}^\nu \psi_i.
\end{eqnarray}
Therefore, for the fermion tensors $J_{d}$, $J_{u}$, and $J_{Q,1}$, one can easily see that: 
\begin{eqnarray}
({\cal CP})  J^{\mu\nu}_{d,pr}(t,\mathbf{r}) ({\cal CP})^\dagger&&=(J_{d,rp})_{\mu\nu}(t,-\mathbf{r}),\\
({\cal CP})  J^{\mu\nu}_{u,pr}(t,\mathbf{r}) ({\cal CP})^\dagger&&=(J_{u,rp})_{\mu\nu}(t,-\mathbf{r}),\\
({\cal CP})  J^{\mu\nu}_{(Q,1),pr}(t,\mathbf{r}) ({\cal CP})^\dagger&&=(J_{(Q,1),rp})_{\mu\nu}(t,-\mathbf{r})
\end{eqnarray}
while for the $J^I_{(Q,3)\mu\nu}$, we have:
\begin{eqnarray}
({\cal CP})J^{I\mu\nu}_{(Q,3),pr}({\cal CP})^\dagger &&= ({\cal CP}) i\tau^{I}_{ji}\bar{q}_{{\rm L}p}^j\gamma^\mu \overleftrightarrow{D}^\nu q_{{\rm L}ri}({\cal CP})^\dagger\nonumber \\
 &&=i\tau^{I}_{ji}\bar{q}_{{\rm L}ri}\gamma^\mu \overleftrightarrow{D}^\nu q_{{\rm L}pj}\nonumber \\
  &&=is^I\tau^{I}_{ij}\bar{q}_{{\rm L}ri}\gamma^\mu \overleftrightarrow{D}^\nu q_{{\rm L}pj}=s^I(J^I_{(Q,3),rp})_{\mu\nu}.
\end{eqnarray}

With the aforementioned transformation properties, we can find that, for the flavor diagonal components, only three operators ${\cal O}_5$, ${\cal O}_7$ and ${\cal O}_{11}$ are CP violating, operators ${\cal O}_{12}$ and ${\cal O}_{13}$ even though containing $\tilde{W}$ are still CP conserving, because under the transformation the fermion tensor $J^{I\mu\nu}_{(Q,3)}$ does not contribute additional minus sign as $W^I$ in does eq.~\eqref{eq:CPWWWt}.

\section{Differential Cross-Section in the Non-COM Frame}\label{app:dsignoncom}

The differential cross-section for the 2-to-2 process can be expressed in the following form:
\begin{eqnarray}
d\sigma_{1+2\to 3+4} = \frac{1}{4F(2\pi)^2} \frac{d^3p_3}{2E_3} \frac{d^3p_4}{2E_4}\delta^{(4)}(p_3+p_4-p_1-p_2)\left|{\cal M}\right|^{2}_{1+2\to 3+4},\label{eq:dsigma}
\end{eqnarray}
where $F$ is the Lorentz invariant M\o ller flux factor defined by:
\begin{eqnarray}
F=E_1E_2v_{12}=\sqrt{(p_1\cdot p_2)^2-m_1^2m_2^2},
\end{eqnarray}
such that the whole formula in eq~\eqref{eq:dsigma} is Lorentz invariant. However, the partially integrated differential cross-section, e.g. the differential cross-section with respect to the solid angle $d\sigma/d\Omega$, may not be Lorentz invariant as the integration can break the Lorentz invariance. For the pp collider, the COM frame of the parton system generally is not the COM frame of the proton system. Since the lab frame coincides with the COM of the proton system, it is convenient to derive a formula for the differential cross-section with respect to the solid angle in the COM frame of the proton system -- $d\sigma/d\Omega^*$  as follows.
Since the amplitude square $\left|{\cal M}\right|^{2}$ and the flux factor $F$ are Lorentz invariant, we focus on the simplification of the phase space and the $\delta$-function:
\begin{eqnarray}
 &&\frac{d^3p_3}{2E_3} \frac{d^3p_4}{2E_4}\delta^{(4)}(p_3+p_4-p_1-p_2)\nonumber \\ =&&d^4p_3\delta(p_3^2-m_3^2)\Theta(E_3)d^4p_3\delta(p_4^2-m_4^2)\Theta(E_4)\delta^{(4)}(p_3+p_4-p_1-p_2)\nonumber \\
 =&&d^4p_3\delta(p_3^2-m_3^2)\Theta(E_3)d^4p_3\delta((p_1+p_2-p_3)^2-m_4^2)\Theta(E_1+E_2-E_3)\nonumber\\
 =&&dE_3d\Omega \frac{\sqrt{E_3^2-m_3^2}}{2}\delta(S-2(p_1+p_2)\cdot p_3+m_3^2-m_4^2)\Theta(E_1+E_2-E_3),\label{eq:dsigsimp1}
\end{eqnarray}
where $S$ is the COM energy square of the parton system. Up to now we have not specified the reference frame yet, here we assume that the two initial partons are massless and their momenta point along with the z-axis in the COM frame of proton, thus their 4-momentum can be expressed as:
\begin{eqnarray}
 p_1=(x_1\sqrt{s}/2,0,0,x_1\sqrt{s}/2),\quad p_2=(x_2\sqrt{s}/2,0,0,-x_2\sqrt{s}/2),
\end{eqnarray}
where $\sqrt{s}$ is the center of mass energy of the proton system, and $x_1$ and $x_2$ are the factions of the proton energy that are carried by the corresponding partons, and indeed the COM energy of the parton system can be expressed as $\sqrt{S}=\sqrt{x_1x_2 s}$. 
Suppose the momentum of the final state particle 3 pointing in the direction with polar angle $\theta$ with respect to the z-axis, then the $\delta$-function in the last line in eq.~\eqref{eq:dsigsimp1} fixes the dependence of $E_3$ on the scattering angle $\theta$ by the following formula:
\begin{eqnarray}\label{eq:eqforE3}
 x_1x_2 s+m_3^2-m_4^2+(x_1-x_2)\sqrt{s}\cos\theta\sqrt{E_3^2-m_3^2}-(x_1+x_2)\sqrt{s}E_3=0,
\end{eqnarray}
which gives the solution to $E_3$:
\begin{eqnarray}
&&E_3(x_1,x_2,\theta)=\nonumber \\
 &&\Bigl[\left(x_1-x_2\right) \cos (\theta ) \sqrt{ \left(m_3^2 s \left(x_1-x_2\right){}^2 \cos ^2(\theta )-m_3^2 \left(2 m_4^2+s \left(x_1^2+x_2^2\right)\right)+\left(m_4^2-s x_1 x_2\right){}^2+m_3^4\right)}\nonumber \\
&& +
   \left(x_1+x_2\right) \left(m_3^2-m_4^2+s x_1 x_2\right)\Bigr]/\left[\sqrt{s} \left(\left(x_1+x_2\right){}^2-\left(x_1-x_2\right){}^2 \cos ^2(\theta )\right)\right].\label{eq:solE3}
\end{eqnarray}
One can easily verify that the above result reproduces expression in the center of mass frame of parton by setting $x_1=x_2$:
\begin{eqnarray}
E_3^{\rm COM} = \frac{m_3^2-m_4^2+S x_1^2s}{2\sqrt{s} x_1}=\frac{m_3^2-m_4^2+S}{2S},
\end{eqnarray}
where we have used the relation $S=x_1^2s$ when $x_1=x_2$. 
Therefore, using eq.~\eqref{eq:dsigsimp1} and the solution~\eqref{eq:solE3} one can define the differential cross-section in the pp COM reference frame as:
\begin{eqnarray}
\frac{d\sigma_{1+2\to 3+4}}{d\Omega}(x_1,x_2,\theta) =\frac{1}{4x_1x_2 s^{3/2}(2\pi)^2}\frac{|\mathbf{p}_3|^2}{\left|(x_1-x_2)E_3\cos\theta-(x_1+x_2)|\mathbf{p}_3|\right|}\left|{\cal M}\right|^{2}_{1+2\to 3+4},\nonumber \\\label{eq:dsigsimp2}
\end{eqnarray}
where $|\mathbf{p}_3|=\sqrt{E_3^2-m_3^2}$ with $E_3$ to be the solution in eq.~\eqref{eq:solE3}, and we also set the $F= {x_1x_2s}/2$ for massless initial partons.
Furthermore, the existence of the solution of $E_3$ will put a constraint on the $x_1$ vs $x_2$ plane when fixing the scattering angle $\theta$ of the particle 3 ($W^+$ in our process), which is summarized in the following inequalities for $\theta\neq \pi/2$:
\begin{eqnarray}
&&\left(m_3^2 s \left(x_1-x_2\right){}^2 \cos ^2(\theta )-m_3^2 \left(2 m_4^2+s \left(x_1^2+x_2^2\right)\right)+\left(m_4^2-s x_1 x_2\right){}^2+m_3^4\right)>0,\\
&& x_1x_2 s+m_3^2-m_4^2+(x_1-x_2)\sqrt{s}\sqrt{E_3^2-m_3^2}\cos\theta>0,\\
&&\left[(x_1+x_2)\sqrt{s}E_3-(x_1x_2 s+m_3^2-m_4^2)\right]/(x_1-x_2)\sqrt{s}\cos\theta>0,
\end{eqnarray}
where the first inequality comes from the requirement of expression under the square root to be larger than zero, the second and the third ones are obtained by demanding $E_3$ and $\sqrt{E_3^2-m3^2}$ larger than zero. For $\theta=\pi/2$, the eq.~\eqref{eq:eqforE3} becomes a linear equation, so the only requirement is $E_3>m_3$.

We can now use eq.~\eqref{eq:dsigsimp2} to derive the hadronic differential cross-section with respect to the solid angle in the pp COM frame:
\begin{eqnarray}
\frac{d\sigma_{p+p\to 3+4}}{d\Omega} = &&\sum_{q_1,q_2}\int dx_1dx_2\ \left[f_{q_1}(x_1)f_{q_2}(x_2)\frac{d\sigma_{1+2\to 3+4}}{d\Omega}(x_1,x_2,\theta)\right.\nonumber \\
&&\quad\quad \left. +f_{q_2}(x_1)f_{q_1}(x_2)\frac{d\sigma_{2+1\to 3+4}}{d\Omega}(x_1,x_2,\theta)\right]/(1+\delta_{q_1q_2})\nonumber \\
=&& \sum_{q_1,q_2}\int dx_1dx_2 \frac{1}{2x_1x_2 s^{3/2}(2\pi^2)}\frac{|\mathbf{p}_3|^2}{\left|(x_1-x_2)E_3\cos\theta-(x_1+x_2)|\mathbf{p}_3|\right|}\nonumber \\
&&\left[f_{q_1}(x_1)f_{q_2}(x_2)\left|{\cal M}\right|^{2}_{1+2\to 3+4}+f_{q_2}(x_1)f_{q_1}(x_2)\left|{\cal M}\right|^{2}_{2+1\to 3+4} \right]/(1+\delta_{q_1q_2}),\nonumber \\
\end{eqnarray}
where $\delta_{q_1q_2}$ equals $1$ for $q_1=q_2$ and zero otherwise, and the invariant scattering amplitudes $\left|{\cal M}\right|^{2}_{2+1\to 3+4}$ and $\left|{\cal M}\right|^{2}_{1+2\to 3+4}$ are related by exchange of the momenta $p_1$ and $p_2$, which yields following equations:
\begin{eqnarray}
\left|{\cal M}\right|^{2}_{1+2\to 3+4}\equiv \left|{\cal M}\right|^{2}_{1+2\to 3+4}(S,T),\\
\left|{\cal M}\right|^{2}_{2+1\to 3+4}=\left|{\cal M}\right|^{2}_{1+2\to 3+4}(S,U),
\end{eqnarray}
where $S,T,U$ are Mandelstam variables for the parton system defined by:
\begin{eqnarray}
S=(p_1+p_2)^2=(p_3+p_4)^2,\\
T=(p_1-p_3)^2=(p_2-p_4)^2,\\
U=(p_1-p_4)^2=(p_2-p_3)^2.
\end{eqnarray}
If one uses the pseudo-rapidity $\eta$ as the angular variable instead of the polar angle, then the differential cross-section with respect to the pseudo-rapidity $\eta$ is related to the differential cross-section of the polar angle $\theta$ in the following formulas:
\begin{eqnarray}
&&\eta=-\ln(\tan\frac{\theta}{2}),\\
&&\frac{d\sigma}{d\eta} = \frac{d\sigma}{d\theta}\frac{d\theta}{d\eta}=\frac{d\sigma}{d\theta}\times\sin\theta.
\end{eqnarray}

\bibliographystyle{JHEP}
\bibliography{dim8diboson}

\end{document}